\documentclass[sigconf]{acmart}

\def\BibTeX{{\rm B\kern-.05em{\sc i\kern-.025em b}\kern-.08emT\kern-.1667em\lower.7ex\hbox{E}\kern-.125emX}}

\usepackage{booktabs} 

\copyrightyear{2019} 
\acmYear{2019} 
\setcopyright{acmcopyright}
\acmConference[ESEC/FSE '19]{Proceedings of the 27th ACM Joint European Software Engineering Conference and Symposium on the Foundations of Software Engineering}{August 26--30, 2019}{Tallinn, Estonia}
\acmBooktitle{Proceedings of the 27th ACM Joint European Software Engineering Conference and Symposium on the Foundations of Software Engineering (ESEC/FSE '19), August 26--30, 2019, Tallinn, Estonia}
\acmPrice{15.00}
\acmDOI{10.1145/3338906.3338935}
\acmISBN{978-1-4503-5572-8/19/08}

\usepackage{booktabs}
\usepackage{moreverb}
\usepackage{fontenc}
\usepackage{amsmath}
\usepackage{amssymb}
\usepackage{fancybox}
\usepackage{color}
\usepackage{colortbl}
\usepackage{array}
\usepackage{multirow}
\usepackage{multicol}
\usepackage{listings}
\usepackage{makecell}
\usepackage{graphicx}
\usepackage{setspace}
\usepackage{footmisc}
\usepackage{wasysym}
\usepackage[htt]{hyphenat}

\usepackage[ruled,vlined,lined,commentsnumbered,linesnumbered]{algorithm2e}
\usepackage{balance}
\usepackage{csvsimple}
\usepackage{longtable}
\usepackage{url}
\usepackage{lscape}
\usepackage{rotating}
\usepackage{tikz}
\usepackage[normalem]{ulem}
\usepackage{enumitem}

\usepackage[most]{tcolorbox}

\usepackage{threeparttable}

\usepackage{marvosym}
\usepackage{pifont}
\usepackage{tabularx}
\usepackage{listings}

\newcolumntype{L}[1]{>{\raggedright\let\newline\\\arraybackslash\hspace{0pt}}m{#1}}
\newcolumntype{C}[1]{>{\centering\let\newline\\\arraybackslash\hspace{0pt}}m{#1}}
\newcolumntype{R}[1]{>{\raggedleft\let\newline\\\arraybackslash\hspace{0pt}}m{#1}}

\definecolor{codegreen}{rgb}{0,0.6,0}
\definecolor{codegray}{rgb}{0.5,0.5,0.5}
\definecolor{codepurple}{rgb}{0.58,0,0.82}
\definecolor{codered}{rgb}{1,0.6,0.6}
\definecolor{backcolour}{rgb}{0.95,0.95,0.92}
\definecolor{lightgray}{gray}{0.9}

\lstdefinestyle{mystyle}{
    commentstyle=\color{codegreen},
    keywordstyle=\color{magenta},
    numberstyle=\tiny\color{codegray},
    stringstyle=\color{codepurple},
    basicstyle=\footnotesize,
    breakatwhitespace=false,
    breaklines=true,
    captionpos=b,
    keepspaces=true,
    showspaces=false,
    showstringspaces=false,
    showtabs=false,
    tabsize=2
}

\lstdefinelanguage{diff}{
}

\setlist{noitemsep} 

\definecolor{darkpastelred}{rgb}{0.76, 0.23, 0.13}
\definecolor{ao(english)}{rgb}{0.0, 0.5, 0.0}

\definecolor{darkpastelred}{rgb}{0.76, 0.23, 0.13}
\definecolor{ao(english)}{rgb}{0.0, 0.5, 0.0}

\definecolor{yellow}{RGB}{255,255,153}
\definecolor{grey}{RGB}{224,224,224}

\newboolean{showcomments}
\setboolean{showcomments}{true}
\ifthenelse{\boolean{showcomments}}
 { \newcommand{\mynote}[2]{
      \fbox{\bfseries\sffamily\scriptsize#1}
        {\small$\blacktriangleright$\textsf{\emph{#2}}$\blacktriangleleft$}}}
        { \newcommand{\mynote}[2]{}}

\definecolor{DarkOrange}{rgb}{0.8,0.3,0.0}
\definecolor{DarkCyan}{rgb}{0.0, 0.55, 0.55}

\newcolumntype{?}{!{\vrule width 1pt}}

\newcommand{\toolname}{\texttt{iFixR}\xspace}

\newcommand{\fixpattern}[1]{
\vspace{-0.4cm}
\begin{tcolorbox}[tile,size=fbox,boxsep=0mm,boxrule=0pt,top=0pt,bottom=0pt,
borderline west={0mm}{0pt}{black!5!white},colback=black!5!white] 
\em #1
\end{tcolorbox}
}

\newcommand{\constraint}[1]{
\begin{tcolorbox}[tile,size=fbox,boxsep=0.5mm,boxrule=0pt,top=0pt,bottom=0pt,
borderline west={0.5mm}{0pt}{black!5!white},colback=black!5!white]
\em #1
\end{tcolorbox}
}

\newcommand{\find}[1]{
\begin{tcolorbox}[tile,size=fbox,boxsep=0.5mm,boxrule=0pt,top=0pt,bottom=0pt,
borderline west={0.5mm}{0pt}{blue!5!white},colback=blue!5!white]
\em #1
\end{tcolorbox}
}

\newcount\mytempcount

\begin{document}
\title{{iFixR}: Bug Report driven Program Repair}

\author{Anil Koyuncu$^{1}$, Kui Liu$^{1, \ast}$, Tegawend\'e F. Bissyand\'e$^{1}$, Dongsun Kim$^{1,2}$, Martin Monperrus$^{3}$, Jacques Klein$^{1}$, and Yves Le Traon$^{1}$}

    \affiliation{\institution{$^{1}$University of Luxembourg, Luxembourg, \{anil.koyuncu, kui.liu,tegawende.bissyande,jacques.klein, yves.letraon\}@uni.lu}}
    \affiliation{\institution{$^{2}$Furiosa.ai, Republic of Korea, darkrsw@furiosa.ai}}
    \affiliation{\institution{$^{3}$KTH Royal Institute of Technology, Sweden, martin.monperrus@csc.kth.se}}

\thanks{$^\ast$Corresponding author, the same contribution as the first author.}

\renewcommand{\shortauthors}{A. Koyuncu et al.}

\begin{abstract}
Issue tracking systems are commonly used in modern software development for collecting feedback from users and developers. An ultimate automation target of software maintenance is then the systematization of patch generation for user-reported bugs. Although this ambition is aligned with the momentum of automated program repair, the literature has, so far, mostly focused on {\em generate-and-validate} setups where fault localization and patch generation are driven by a well-defined test suite.
On the one hand, however, the common (yet strong) assumption on the existence of relevant test cases does not hold in practice for most development settings: many bugs are reported without the available test suite being able to reveal them. On the other hand, for many projects, the number of bug reports generally outstrips the resources available to triage them.
Towards increasing the adoption of patch generation tools by practitioners, we investigate a new repair pipeline, \toolname, driven by bug reports: (1) bug reports are fed to an IR-based fault localizer; (2) patches are generated from fix patterns and validated via regression testing; (3) a prioritized list of generated patches is proposed to developers.
We evaluate \toolname on the Defects4J dataset, which we enriched (i.e., faults are linked to bug reports) and carefully-reorganized (i.e., the timeline of test-cases is naturally split).
\toolname generates genuine/plausible patches for 21/44 Defects4J faults with its IR-based fault localizer.
\toolname accurately places a genuine/plausible patch among its top-5 recommendation for 8/13 of these faults (without using future test cases in generation-and-validation).

\end{abstract}

\ccsdesc[500]{Software and its engineering~Software verification and validation}
\ccsdesc[100]{Software and its engineering~Software testing and debugging}

\keywords{Information retrieval, fault localization, automatic patch generation.}

\maketitle

\section{Introduction}

Automated program repair (APR) has gained incredible momentum in the last decade.
Since the seminal work by Weimer et al.~\cite{westley2009automatically} who relied on genetic programming to evolve program variants until one variant is found to satisfy the functional constraints of a test suite, the community has been interested in test-based techniques to   repair {\em programs without specifications}.
Thus, various approaches~\cite{nguyen2013semfix,westley2009automatically,le2012genprog,kim2013automatic,coker2013program,ke2015repairing,mechtaev2015directfix,long2015staged,le2016enhancing,long2016automatic,chen2017contract,le2017s3,long2017automatic,xuan2017nopol,xiong2017precise,jiang2018shaping,wen2018context,hua2018towards,liu2019avatar,liu2019you, liu2019learning} have been proposed in the literature aiming at reducing manual debugging efforts through automatically generating patches.
Beyond fixing {\em syntactic errors}, i.e., cases where the code violates some programming language specifications~\cite{gupta2017deepfix}, the current challenges lie in fixing {\em semantic bugs}, i.e., cases where implementation of program behavior deviates from developer's intention~\cite{mechtaev2018semantic, just2018comparing}.

Ten years ago, the work of Weimer et al.~\cite{westley2009automatically} was explicitly motivated by the fact that, despite significant advances in
specification mining (e.g.,~\cite{le2009specification}), formal specifications are rarely available. Thus, test suites represented an affordable approximation to program specifications. Unfortunately, the assumption that {\em test cases are readily available} still does not hold in practice~\cite{kochhar2013empirical,beller2015and, petric2018effectively}. Therefore, while current test-based APR approaches would be suitable in a test-driven development
setting~\cite{beck2003test}, their adoption by practitioners faces a simple reality: developers majoritarily (1) write few tests~\cite{kochhar2013empirical}, (2) write tests after the source code~\cite{beller2015and}, and (3) write tests to validate that bugs are indeed fixed and will not reoccur~\cite{juristo2006guest}.

Although APR bots~\cite{urli2018design} can come in handy in a continuous integration environment, the reality is that {\em bug reports} remain the main source of the stream of bugs that developers struggle to handle daily~\cite{anvik2006should}. Bugs are indeed reported in natural language, where users tentatively describe the execution scenario that was being carried out and the unexpected outcome (e.g., crash stack traces). Such bug reports constitute an essential artifact within a software development cycle and can become an overwhelming concern for maintainers. For example, as early as in 2005, a triager of the Mozilla project was reported in~\citep[page~363]{anvik2006should} to have commented that:

\renewenvironment{quote}{%
   \list{}{%
     \leftmargin0.5cm   
     \rightmargin\leftmargin
   }
   \item\relax
}
{\endlist}

\begin{quote}
	``{\em Everyday, almost 300 bugs appear that need triaging. This is far too much for only the Mozilla programmers to handle.}''
\end{quote}

\begin{figure}[!t]
	\includegraphics[width=\linewidth]{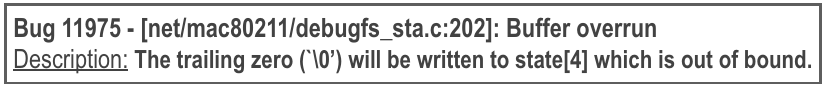}
	\caption{Example of Linux bug report addressed by R2Fix.}
	\label{fig:r2fix}
\end{figure}

However, very few studies~\cite{liu2013r2fix,bissyande2015harvesting} have undertaken to automate patch generation based on bug reports. To the best of our knowledge, Liu et al.~\cite{liu2013r2fix} proposed the most advanced study in this direction.
Unfortunately, their R2Fix approach carries several caveats: as illustrated in Figure~\ref{fig:r2fix}, it focuses on perfect bug reports~\citep[page~283]{liu2013r2fix} (1) which explicitly include localization information, (2) where the symptom (e.g., buffer overrun) is explicitly indicated by the reporter, and (3) which are about one of the following three simple bug types: Buffer overflow, Null Pointer dereference or memory leak. R2Fix runs a straightforward classification to identify the bug category and uses a match and transform engine (e.g., Coccinelle~\cite{padioleau2008documenting}) to generate patches. 
As the authors admitted, their target space represents \textless1\% of bug reports in their dataset.
Furthermore, it should be noted that, given the limited scope of the changes implemented in its fix patterns, R2Fix does not need to run tests for verifying that the generated patches do not break any functionality.

{\bf This paper.} We propose to investigate the feasibility of a program repair system driven by bug reports, thus we replace classical spectrum-based fault localization with Information Retrieval (IR)-based fault localization. Eventually, we propose \toolname, a new program repair workflow which considers a practical repair setup by imitating the fundamental steps of manual debugging.
\toolname works under the following constraint:
\constraint{When a bug report is submitted to the issue tracking system, a relevant test case reproducing the bug may not be readily available.}

Therefore, \toolname is leveraged in this study to assess {\em to what extent an APR pipeline is feasible under the practical constraint of limited test suites}. \toolname uses bug reports written in natural language as the main input.
Eventually, we make the following contributions:
\begin{itemize}[leftmargin=*]
	\item We present the architecture of a program repair system adapted to the constraints of maintainers dealing with user-reported bugs. In particular, \toolname replaces classical spectrum-based fault localization with Information Retrieval (IR)-based fault localization.
	\item We propose a strategy to prioritize patches for recommendation to developers. Indeed, given that we assume only the presence of regression test cases to validate patch candidates, many of these patches may fail on the future test cases that are relevant to the reported bugs. We order patches to present correct patches first.
	\item We assess and discuss the performance of \toolname on the Defects4J benchmark to compare with the state-of-the-art APR tools. To that end, we provide a refined Defects4J benchmark for APR targeting bug reports. Bugs are  carefully linked with the corresponding bug reports, and for each bug we are able to dissociate {\em future test cases} that were introduced after the relevant fixes.
	\end{itemize}
	
Overall, experimental results show that there are promising research directions to further investigate towards the integration of automatic patch generation in actual software development cycles. In particular, our findings suggest that IR-based fault localization errors lead less to overfitting patches than spectrum-based fault localization errors. Furthermore, \toolname offers provides comparable results to most state-of-the-art APR tools, although it is run under the constraint that post-fix knowledge (i.e., future test cases) is not available. Finally, \toolname's prioritization strategy tends to place more correct/plausible patches on top of the recommendation list.

\section{Motivation}
\label{sec:motivation}
We motivate our work by revisiting two essential steps in APR:
\begin{enumerate}[leftmargin=*]
	\item During {\em fault localization}, relevant program entities are identified as suspicious locations that must be changed. Commonly, state-of-the-art APR tools leverage spectrum-based fault localization (SBFL)~\cite{nguyen2013semfix, westley2009automatically,claire2012genprog,ke2015repairing,mechtaev2015directfix,long2015staged,le2016enhancing,chen2017contract, le2017s3,long2017automatic,xuan2017nopol,xiong2017precise,liu2019tbar}, which uses execution coverage information of passing and failing test cases to predict buggy statements. We dissect the construction of the Defects4J dataset to highlight the practical challenges of fault localization for user-reported bugs.
	\item Once a patch candidate is generated, the {\em patch validation} step ensures that it is actually relevant for repairing the program. Currently, widespread test-based APR techniques use test suites as the repair oracle. This however is challenged by the incompleteness of test suites, and may further not be inline with developer requirements/expectations in the repair process.
\end{enumerate}
%

\subsection{Fault Localization Challenges}
Defects4J is a manual curated dataset widely used in the APR literature~\cite{xin2017leveraging,chen2017contract,saha2017elixir,xiong2018identifying,wen2018context,hua2018towards}.
Since Defects4J was not initially built for APR, the real order of precedence between the bug report, the patch and the test case is being overlooked by the dataset users. Indeed, Defects4J offers a user-friendly way of checking out buggy versions of programs with all relevant test cases for readily benchmarking test-based systems.
We propose to carefully examine the actual bug fix commits associated with Defects4J bugs and study how the test suite is evolved. Table~\ref{tab:defects4j_tests} provides detailed information.

\begin{table}[!h]
	\centering
	\scriptsize
	\caption{Test case changes in fix commits of Defects4J bugs.}
	\begin{threeparttable}
		\begin{tabular}{lC{1.3cm}}
		\toprule
		{\bf Test case related commits}	& {\bf\# bugs}\\
		\hline
		Commit does not alter test cases & 14\\
		Commit is inserting new test case(s) and updating previous test case(s) & 62\\
		Commit is updating previous test case(s) (without inserting new test cases) & 76\\
		Commit is inserting new test case(s) (without updating previous test cases) & 243\\
		\bottomrule
		\end{tabular}
	\end{threeparttable}
	\label{tab:defects4j_tests}
\end{table}

Overall, for 96\%(=381/395) bugs, the relevant test cases are actually {\em future data} with respect to the bug discovery process.
This finding suggests that, in practice, even the fault localization  may be challenged in the case of user-reported bugs, given the lack of relevant test cases. The statistics listed in Table~\ref{tab:statistics2} indeed shows that if future test cases are dropped, no test case is failing when executing buggy program versions for 365 (i.e., 92\%) bugs.

 \begin{table}[!h]
	\centering
	\caption{Failing test cases after removing future test cases.}
	\scriptsize
	\label{tab:statistics2}
	\resizebox{\linewidth}{!}{
	\begin{threeparttable}
		\begin{tabularx}{\linewidth}{Xc}
			\toprule
			{\bf Failing test cases} & {\bf\# bugs}\\
			\hline
			Failing test cases exist (and no future test cases are committed) & 14 \\
			Failing test cases exist  (but future test cases update the test scenarios) & 9\\
			Failing test cases exist (but they are fewer when considering future test cases) & 4 \\
			Failing test cases exist (but they differ from future test cases which trigger the bug) & 3\\
			No failing test case exists (i.e., only future test cases trigger the bug)& 365\\
			\bottomrule
		\end{tabularx}
	\end{threeparttable}
	}
\end{table}

In the APR literature, fault localization is generally performed using the GZoltar~\cite{campos2012gzoltar} testing framework and a SBFL formula~\cite{wong2016survey} (e.g., Ochiai~\cite{abreu2007accuracy}).
To support our discussions, we attempt to perform fault localization without the future test cases to evaluate the performance gap. Experimental results (see details forward in Table~\ref{tab:gzoltar} of Section~\ref{sec:results}) expectedly reveal that the majority of the Defects4J bugs (i.e., 375/395) cannot be localized by SBFL at the time the bug is reported by users.
 
\constraint{It is necessary to investigate alternate fault localization approaches that build on bug report information since relevant test cases are often unavailable when users report bugs.}

\begin{figure*}[!ht]
 	\includegraphics[width=0.8\textwidth]{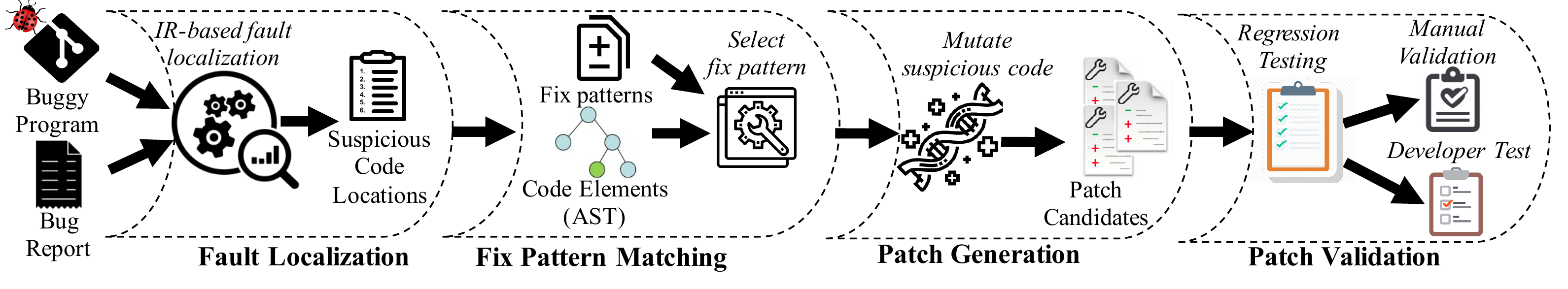}
	\caption{The \toolname Program Repair Workflow.}
	\label{fig:repairPipeline}
\end{figure*}

\subsection{Patch Validation in Practice}

The repair community has started to reflect on the {\em acceptability}~\cite{kim2013automatic,monperrus2014critical} and {\em correctness}~\cite{smith2015cure,xiong2018identifying} of the patches generated by APR tools. Notably, various studies~\cite{smith2015cure,qi2015analysis,yang2017better,bohme2017bug,le2018overfitting} have raised concerns about overfitting patches: a typical APR tool that uses a test suite as the correctness criterion can produce a patched program that actually overfits the test-suite (i.e., the patch makes the program pass all test cases but does not actually repair it).
Recently, new research directions~\cite{yu2017test,xin2017identifying} are being explored in the automation of test case generation for APR to overcome the overfitting issue. Nevertheless, so far they have had minimal positive impact due to the oracle problem~\cite{yu2018alleviating} in automatic test generation.

At the same time, the software industry takes a more systematic approach for patch validation by developers. For instance, in the open-source community, the Linux development project has integrated a patch generation  engine to automate collateral evolutions that are validated by maintainers~\cite{padioleau2008documenting,koyuncu2017impact}.
In proprietary settings, Facebook has recently reported on their {\em Getafix}~\cite{getafix2019} tool,
which automatically suggests fixes to their developers. Similarly, Ubisoft developed {\em Clever}~\cite{nayrolles2018clever} to detect risky commits at commit-time using patterns of programming mistakes from the code history.

\constraint{Patch recommendation for validation by developers is acceptable in the software development communities. It may thus be worthwhile to focus on tractable techniques for recommending patches in the road to fully automated program repair.}

\section{The \toolname Approach}
\label{sec:approach}

Figure~\ref{fig:repairPipeline} overviews the workflow of the proposed \toolname approach.
Given a defective program, we consider the following issues:

\begin{enumerate}[leftmargin=*]
	\item {\bf Where is the bug?} We take as input the bug report in natural language submitted by the program user. We rely on the information in this report to localize the bug positions.
	\item {\bf How should we change the code?} We apply fix patterns that are recurrently found in real-world bug fixes. Fix patterns are selected following the structure of the abstract syntax tree representing the code entity of the identified suspicious code.
	\item {\bf Which patches are valid?} We make no assumptions on the availability of {\em positive test cases}~\cite{westley2009automatically} that encode functionality requirements at the time the bug is discovered. Nevertheless, we leverage existing test cases to ensure, at least, that the patch does not regress the program.
	\item {\bf Which patches do we recommend first?} In the absence of a complete test suite, we cannot guarantee that all patches that pass regression tests will fix the bug. We rely on heuristics to re-prioritize the validated patches in order to increase the probability of placing a correct patch on top of the list.
\end{enumerate}

\subsection{Input: Bug reports}

Issue tracking systems (e.g., Jira) are widely used by software development communities in the open source and commercial realms.
Although they can be used by developers to keep track of the bugs that they encounter and the features to be implemented, issue tracking systems allow for user participation as a communication channel for collecting feedback on software executions in production.

Table~\ref{tab:sampleBR2} illustrates a typical bug report when a user of the LANG library code has encountered an issue while using the {\em NumberUtils} API. A description of erroneous behavior is provided. Occasionally, the user may include in the bug description some information on how to reproduce the bug. Oftentimes, users simply insert code snippets or dump the execution stack traces.

In this study, among our dataset of 162 bug reports, we note that only 27 (i.e., $\sim$17\%) are reported by users who are also developers\footnote{We rely on email addresses of committers and issue reporters to intersect users and developers} contributing to the projects. 15 (i.e., $~\sim$9\%) bugs are reported and again fixed by the same project contributors. These percentages suggest that, for the majority of cases, the bug reports are indeed genuinely submitted by users of the software who require project developers' attention.



\begin{table}[!h]
	\centering
	\scriptsize
	\caption{Example bug report (Defects4J Lang-7).}

		\begin{tabularx}{\linewidth}{l|X}
		\toprule
		Issue No. 	& LANG-822\\
		\hline
		Summary & NumberUtils\#createNumber - bad behaviour for leading "--" \\
		\hline
		Description 
		& {NumberUtils\#createNumber checks for a leading "--" in the string, and returns null if found. This is documented as a work round for a bug in BigDecimal.

Returning nulll is contrary to the Javadoc and the behaviour for other methods which would throw NumberFormatException.

It's not clear whether the BigDecimal problem still exists with recent versions of Java. However, if it does exist, then the check needs to be done for all invocations of BigDecimal, i.e. needs to be moved to createBigDecimal.
		 }\\
		\bottomrule
		\end{tabularx}
	\label{tab:sampleBR2}
\end{table}

Given the buggy program version and a bug report, \toolname must unfold the workflow for precisely identifying (at the statement level) the buggy code locations.
We remind the reader that, in this step, future test cases cannot be relied upon.
We consider that if such test cases could have triggered the bug, a continuous integration system would have helped developers deal with the bug before the software is shipped towards users.



\subsection{Fault Localization w/o Test Cases}
To identify buggy code locations within the source code of a program, we resort to Information Retrieval
(IR)-based fault localization (IRFL)~\cite{parnin2011automated,wang2015evaluating}.
The general objective is to leverage potential similarity between the terms used in a bug report and the source code to identify relevant buggy code locations.
The literature includes a large body of work on IRFL~\cite{zhou2012should,wen2016locus,youm2017improved,wong2014boosting,saha2013improving,wang2014version,lukins2010bug} where researchers systematically extract tokens from a given bug report to formulate a {\em query} to be matched in a search space of {\em documents} formed by the collections of source code files and indexed through tokens extracted from source code.
IRFL approaches then rank the documents based on a probability of relevance (often measured as a similarity score). Highly ranked files are predicted to be the ones that are likely to contain the buggy code.

Despite recurring interest in the literature, with numerous approaches continuously claiming new performance improvements over the state-of-the-art, we are not aware of any adoption in program repair research or practice.
We postulate that one of the reasons is that IRFL techniques have so far focused on file-level localization, which is too coarse-grained (in comparison to spectrum-based fault localization output). Recently, Locus~\cite{wen2016locus} and BLIA~\cite{youm2017improved} are state-of-the-art techniques which narrow down localization, respectively to the code change or the method level.
Nevertheless, to the best of our knowledge, no IRFL technique has been proposed in the literature for statement-level localization.

 In this work, we develop an algorithm to rank suspicious statements based on the output (i.e., files) of a state-of-the-art IRFL tool, thus yielding a fine-grained IR-based fault localizer which will then be readily integrated into a concrete patch generation pipeline.

\subsubsection{Ranking Suspicious Files}
We leverage an existing IRFL tool. Given that expensive extractions of tokens from a large corpus of bug reports is often necessary to tune IRFL tools~\cite{lee2018bench4bl}, we selected a tool for which the authors provide datasets and pre-processed data.
We use the D\&C~\cite{koyuncu2019d} as the specific implementation of file-level IRFL available online~\cite{dANDc}
, which is a machine learning-based IRFL tool using a similarity matrix of 70-dimension feature vectors (7 features from bug reports and 10 features from source code files): D\&C uses multiple classifier models that are trained each for specific groups of bug reports.
Given a bug report, the different predictions of the different classifiers are merged to yield a single list of suspicious code files.
Our execution of D\&C (Line~2 in Algorithm~\ref{alg:file2stmt}) is tractable given that we only need to preprocess those bug reports that we must localize. Trained classifiers are already available. We ensure that no data leakage is induced (i.e., the classifiers are not trained with bug reports that we want to localize in this work).

%
%

\subsubsection{Ranking Suspicious Statements}
Patch generation requires fine-grained information on code entities that must be changed.
For \toolname, we propose to produce a standard output, as for spectrum-based fault localization, to facilitate integration and reuse of state-of-the-art patch generation techniques.
To start, we build on the conclusions on a recent large-scale study~\cite{liu2018closer} of bug fixes to limit the search space of suspicious locations to the statements that are more error-prone.
After investigating in detail the abstract syntax tree (AST)-based code differences of over 16\,000 real-world patches from Java projects, Liu et al.~\cite{liu2018closer} reported that the following specific AST statement nodes were significantly more prone to be faulty than others: {\tt IfStatements}, {\tt ExpressionStatements}, {\tt FieldDeclarations}, {\tt ReturnStatements} and {\tt VariableDeclarationStatements}.
Lines~7--17 in Algorithm~\ref{alg:file2stmt} detail the process to produce a ranked list of suspicious statements.

\begin{algorithm}[!h]
\scriptsize
    \SetKwInOut{Input}{Input}
    \SetKwInOut{Variables}{Variables}
    \SetKwInOut{Output}{Output}
    \SetKw{Return}{return}
    \SetKwProg{Fn}{Function}{}{end}
    \SetKwFunction{cosSim}{similarity\textsubscript{cosine}}
    \SetKwFunction{calculateWeights}{main}
    \SetKwFunction{features}{bagOfTokens}
    \SetKwFunction{preprocess}{preprocess}
    \SetKwFunction{irfl}{fileLocalizations}
    \SetKwFunction{selectTop}{selectTop}

\DontPrintSemicolon
    \Input{$br$ : a bug report}
    \Input{$irTool$ : IRFL tool}
    \Output{$S_{score}$ : Suspicious Statements with weight scores}

    \Fn{\calculateWeights($br$,$irTool$)}{
        $F$ $\leftarrow$ \irfl($irTool$,$br$)\;
        $F$ $\leftarrow$ \selectTop($F$,$k$) \;
        $c_b$ $\leftarrow$ \features($br$) \tcc*{$c_b$: Bag of Tokens of bug report}
        $c_b'$ $\leftarrow$ \preprocess($c_b$)\tcc*{tokenization,stopword removal, stemming}

        $v_b$ $\leftarrow$ tfIdfVectorizer($c_b'$) \tcc*{$v_b$: Bug report Feature Vector}

    \For{$f$ in $F$ }{
        $S$ $\leftarrow$ parse($f$) \tcc*{$S$: List of statements}
        \For{$s$ in $S$ }{
            $c_s$ $\leftarrow$ \features($s$) \tcc*{$c_s$: Bag of Tokens of statements}
            $c_s'$ $\leftarrow$ \preprocess($c_s$)\;

            $v_s$ $\leftarrow$ tfIdfVectorizer($c_s'$) \tcc*{$v_s$: Statements Feature Vector}
            
            \tcc{Cosine similarity between bug report and statement}
            $sim_{cos}$ $\leftarrow$ \cosSim($v_b$,$v_s$)\;

            $w_{score}$ $\leftarrow$ $sim_{cos}$ $\times$ $f$.score; \tcc*{score: Suspicious Value}

            $W_{score}$.add($s$,$w_{score}$) \;
        }

    }

        $S_{score}$ $\leftarrow$ $W_{score}$.sort() \;
        \Return $S_{score}$ \;

    %
    }



    \caption{Statement-level IR-based Fault Localization.}
    \label{alg:file2stmt}
\end{algorithm}

Algorithm~\ref{alg:file2stmt} describes the process of our fault localization
approach used in \toolname.
Top $k$ files are selected among the returned list of suspicious files of the IRFL along with their computed suspiciousness scores. Then each file is parsed to retain only the relevant error-prone statements from which textual tokens are extracted. The summary and descriptions of the bug report are also analyzed (lexically) to collect all its tokens. Due to the specific nature of stack traces and other code elements which may appear in the bug report, we use regular expressions to detect stack traces and code elements to improve the tokenization process, which is based on punctuations, camel case splitting (e.g., findNumber splits into find, number) as well as snake case splitting (e.g., find\_number splits into find, number). Stop word removal\footnote{Stop words are from the NTLK framework :\url{https://www.nltk.org/}} is then applied before performing stemming (using the PorterStemmer~\cite{karaa2013information}) on all tokens to create homogeneity with the term's root (i.e., by conflating variants of the same term).
Each bag of tokens (for the bug report, and for each statement) is then eventually used to build a feature vector. We use cosine similarity among the vectors to rank the file statements that are relevant to the bug report.

Given that we considered $k$ files, the statements of each having their own similarity score with respect to the bug report, we weight these scores with the suspiciousness score of the associated file.
Eventually, we sort the statements using the weighted scores and produce a ranked list of code locations (i.e., statements in files) to be recommended as candidate fault locations.



\subsection{Fix Pattern-based Patch Generation}
\label{subsec:generation}
A common, and reliable, strategy in automatic program repair is to generate concrete patches based
on fix patterns~\cite{kim2013automatic} (also referred
to as fix templates~\cite{liu2018mining} or program transformation schemas~\cite{hua2018towards}).
Several APR systems~\cite{kim2013automatic,saha2017elixir,durieux2017dynamic,
liu2018mining,hua2018towards,koyuncu2018fixminer,martinez2018ultra,liu2019avatar,liu2019you,liu2018mining2} in the literature implement this strategy by using diverse sets of fix patterns obtained either via manual generation or automatic mining of bug fix datasets.
In this work, we consider the pioneer {\em PAR} system by Kim et al.~\cite{kim2013automatic}. Concretely, we build on {\em kPAR}~\cite{liu2019you}, an open-source Java implementation of {\em PAR} in which we included a diverse set of fix patterns collected from the literature.
Table~\ref{tab:fps} provides an enumeration of fix patterns used in this work. For more implementation details, we refer the reader to our replication package. All tools and data are released as open source to the community to foster further research into these directions. As illustrated in Figure~\ref{fig:fp}, a fix pattern encodes the recipe of change actions that should be applied to mutate a code element.


\begin{table}[!h]
	\centering
	\scriptsize
	\caption{Fix patterns implemented in \toolname.}
	\begin{threeparttable}
		\begin{tabular}{lr|lr}
		 {\bf Pattern description} & {\bf used by$^\ast$}& {\bf Pattern description} & {\bf used by$^\ast$}\\
		\toprule
		Insert Cast Checker & Genesis &Mutate Literal Expression & SimFix\\
		Insert Null Pointer Checker & NPEFix&Mutate Method Invocation & ELIXIR\\
		Insert Range Checker & SOFix & Mutate Operator & jMutRepair\\
		Insert Missed Statement & HDRepair & Mutate Return Statement &  SketchFix\\
		Mutate Conditional Expression & ssFix & Mutate Variable & CapGen\\
		Mutate Data Type & AVATAR& Move Statement(s) & PAR\\
		Remove Statement(s) & FixMiner& & \\
		\bottomrule
		\end{tabular}
		{$^\ast$ We mention only one example tool even when several tools implement it.}
	\end{threeparttable}
	\label{tab:fps}
\end{table}

\begin{figure}[!h]
	\centering
	\vspace{4mm}
	\fixpattern{
\begin{lstlisting}[language=diff,linewidth={\linewidth},basicstyle=\footnotesize\ttfamily]
(@\phantom{x}\textcolor{codegreen}{+}@)  if (exp instanceof T) {
(@\phantom{x}\hspace{6ex}@) ...(T) exp...; ......
(@\phantom{x}\textcolor{codegreen}{+}@)  }
\end{lstlisting}}
	\caption{Illustration of ``Insert Cast Checker'' fix pattern.}
	\label{fig:fp}
\end{figure}

For a given reported bug, once our fault localizer yields its list of suspicious statements,
\toolname iteratively attempts to select fix patterns for each statement.
The selection of fix patterns is conducted in a na\"ive way based on the context information of each suspicious statement (i.e., all nodes in its abstract syntax tree, AST).
Specifically, \toolname parses the code and traverses each node of the suspicious statement AST from its first child node to its last leaf node in a breadth-first strategy (i.e, left-to-right and top-to-bottom).
If a node matches the context a fix pattern (i.e., same AST node types), the fix pattern will be applied to generate patch candidates by mutating the matched code entity following the recipe in the fix pattern. Whether the node matches a fix pattern or not, \toolname keeps traversing its children nodes and searches fix patterns for them to generate patch candidates successively. This process is iteratively performed until leaf nodes are encountered.

\begin{figure}[!t]
	\centering
	\includegraphics[width=1\linewidth]{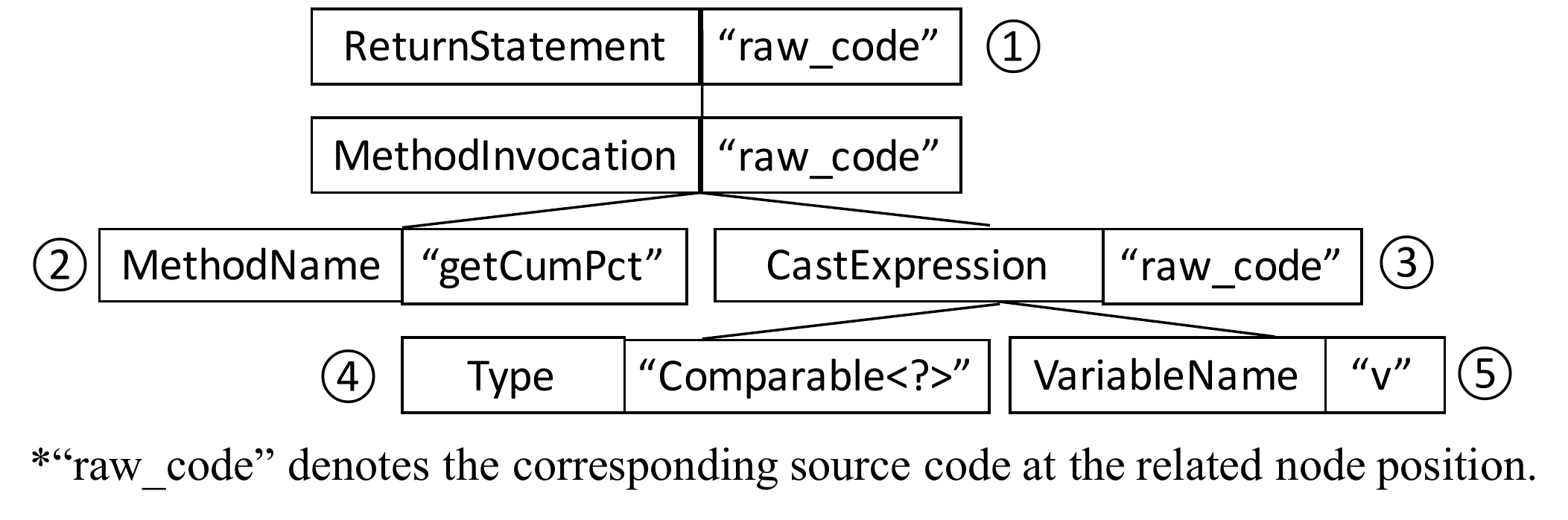}
	\caption{AST of bug Math-75 source code statement.}
	\label{fig:math75AST}
\end{figure}

Consider the example of bug Math-75 illustrated in Figure~\ref{fig:math75}. \toolname parses the buggy statement (i.e., statement at line 302 in the file {\em Frequency.java}) into an AST as illustrated by Figure~\ref{fig:math75AST}.
First, \toolname matches a fix pattern that can mutate the expression in the return statement with other expression(s) returning data of type {\em double}.
It further selects fix patterns for the direct child node (i.e., method invocation: {\tt getCumPct((Comparable<?> v))}) of the return statement.
This method invocation can be matched against fix patterns with two contexts: method name and parameter(s).
With the breadth-first strategy, \toolname assigns a fix pattern, calling another method with the same parameters (cf. PAR~\citep[page~804]{kim2013automatic}), to mutate the method name, and then selects fix patterns to mutate the parameter.
Furthermore, \toolname will match fix patterns for the type and variable of the cast expression respectively and successively.

\begin{figure}[!h]
	\centering
	\vspace{2mm}
\begin{lstlisting}[language=diff,linewidth={\linewidth},frame=tb,basicstyle=\footnotesize\ttfamily]
File: src/main/java/org/apache/commons/math/stat/Frequency.java
Line-301     public double getPct(Object v) {
Line-302(@{\colorbox{codered}{\hspace{8ex} return getCumPct((Comparable<?>) v);\hspace{16ex}}}@)
Line-303     } 
\end{lstlisting}
	\caption{Buggy code of Defects4J bug Math-75.}
	\label{fig:math75}
\end{figure}

\subsection{Patch Validation with Regression Testing}
\label{subsec:validation}
For every reported bug,
fault localization followed by pattern matching and code mutation
will yield a set of patch candidates. In a typical test-based APR system, these patch candidates must let the program pass all test cases (including some {\em positive test cases}~\cite{westley2009automatically},
which encode the actual functional requirements relevant to the bug).
Thus, the patch candidates set is actively pruned to remove
all patches that do not meet these requirements.
In our work, in accordance with our investigation findings that such test cases may not be available at the time the bug is reported (cf. Section~\ref{sec:motivation}), we assume that \toolname cannot reason about {\em future} test cases to select patch candidates.

Instead, we rely only on {\em past} test cases, which were available in the code base,
when the bug is reported. Such test cases are leveraged to perform {\em regression testing}~\cite{yoo2012regression},
which will ensure that,
at least, the selected patches do not obstruct the behavior of
the existing, unchanged part of the software, which is already explicitly encoded by developers in their current test suite.

\subsection{Output: Patch Recommendation List}
Eventually, \toolname produces a ranked recommendation list of patch suggestions for developers.
Until now, the order of patches is influenced mainly by two steps in the workflow:
\begin{enumerate}[leftmargin=*]
	\item localization: our statement-level IRFL yields a ranked list of statements to modify in priority.
	\item pattern matching: the AST node of the buggy code entity is broken down into its children and iteratively navigated in a breadth-first manner to successively produce candidate patches.
\end{enumerate}
Eventually, the produced list of patches has an order, which carries the biases of fault localization~\cite{liu2019you}, and is noised by the pre-set breadth-first strategy for matching fix patterns. We thus design an ordering process with a function\footnote{The domain of the function is a power set $2^{\mathbb{P}}$, and the co-domain ($\mathbb{P}^k$) is a $k$-dimensional vector space~\cite{kolmogorov1999elements} where $k$ is the maximum number of recommended patches, and $\mathbb{P}$ denotes the set of all generated patches.}, $f_{rcmd}: 2^{\mathbb{P}} \rightarrow \mathbb{P}^k$, as follows:%
\begin{equation}
	f_{rcmd} (patches) = (pri_{type} \circ pri_{susp} \circ pri_{change}) (patches)
	\label{eq:pri}
\end{equation}
%
%
where 
$pri_{\ast}$ are three heuristics-based prioritization functions
used in \toolname. $f_{rcmd}$ takes a set of patches validated via regression testing (cf. Section~\ref{subsec:validation}) and
produces an ordered sequence of patches
($f_{rcmd} (patches) = seq_{rcmd} \in \mathbb{P}^k$). We propose the following {\bf heuristics to re-prioritize the patch candidates}:
\begin{enumerate}[leftmargin=*]
	\item {\tt [Minimal changes]}: we favor patches that minimize the differences between the patched program and the buggy program. To that end, patches are ordered following their AST edit script sizes. Formally, we define $pri_{change}: 2^{\mathbb{P}} \rightarrow \mathbb{P}^n$ where
	$n=|patches|$, $pri_{change}(patches) = [p_{i}, p_{i+1}, p_{i+2}, \cdots ]$ and holds {\centering $\forall p \in patches, C_{change}(p_{i}) \leq C_{change}(p_{i+1})$.}  Here, $C_{change}(p)$ is a function that counts the number of deleted and inserted AST nodes by the change actions of $p$.
	\item {\tt [Fault localization suspiciousness]}: when two patch candidates have equal edit script sizes, the tie is broken by using the suspiciousness scores (of the associated statements) yielded during IR-based fault localization. Thus, when $C_{change}(p_{i}) == C_{change}(p_{i+1})$, $pri_{susp}$ re-orders the two patch candidates. We define $pri_{susp}: \mathbb{P}^n \rightarrow \mathbb{P}^n$ such that $pri_{susp}(seq_{change}) = [\cdots, p_{i}, p_{i+1}, \cdots]$ holds $S_{susp}(p_{i}) \geq S_{susp}(p_{i+1})$,
	where \\$seq_{change}$ is the result of $pri_{change}$ and $S_{susp}$ returns a suspicious score of the statement that a given patch $p_i$ changes.
	\item {\tt [Affected code elements]}: after a manual analysis of fix patterns and the performance of associated APR in the literature, we empirically found that some change actions are irrelevant to bug fixing. Thus, for the corresponding pre-defined patterns, \toolname systematically under-prioritizes their generated patches against any other patches, although among themselves the ranking obtained so far (through $pri_{change}$ and $pri_{susp}$) is preserved for those under-prioritized patches. These are patches generated by (i) mutating a literal expression, (ii) mutating a variable into a method invocation or a final static variable, or (iii) inserting a method invocation without parameter. This prioritization, is defined by $pri_{type}: \mathbb{P}^n \rightarrow \mathbb{P}^k$, which returns a sequence of top $k$ ordered patches ($k \leq n=|patches|$).
	To define this prioritization function, we assign natural numbers $j_1,j_2, j_3, j_4 \in \mathbb{N}$ to each patch generation types (i.e., $j_1 \leftarrow$(i), $j_2 \leftarrow$(ii), and $j_3 \leftarrow$(iii), respectively) and  ($j_4 \leftarrow$) everything else, which strictly hold $j_4 > j_1, j_4 > j_2, j_4 > j_3$. This prioritization function takes the result of $pri_{susp}$ and returns another sequence $[p_{i}, p_{i+1}, p_{i+2}, \cdots ]$ that holds $\forall p_{i}, D_{type}(p_{i}) \geq D_{type}(p_{i+1})$.
	$D_{type}$ is defined as $D_{type}:2^{\mathbb{P}} \rightarrow \{j_1,j_2,j_3,j_4\}$ and determines
	how a patch $p_{i}$ has been generated as defined above. From the ordered sequence, the function
	returns the leftmost (i.e., top) $k$ patches as a result.
\end{enumerate}

\section{Experimental Setup}
We now provide details on the experiments that we carry out to assess the \toolname patch generation pipeline for user-reported bugs.
Notably, we discuss the dataset and benchmark, some implementation details before enumerating the research questions.

\subsection{Dataset \& Benchmark}
To evaluate \toolname we propose to rely on the Defects4J~\cite{just2014defects4j} dataset which is widely used as a benchmark in the Java APR literature.
Nevertheless, given that Defects4J does not provide direct links to the bug reports that are associated with the benchmark bugs, we must undertake a {\em fairly accurate} bug linking task~\cite{thomas2013impact}. Furthermore, to realistically evaluate \toolname, we must reorganize the dataset test suites to accurately simulate the context at the time the bug report is submitted by users.

\subsubsection{Bug linking} To identify the bug report describing a given bug in the Defects4J dataset we focus on recovering the links between the bug fix commits and bug reports from the issue tracking system.
Unfortunately, projects Joda-Time, JFreeChart and Closure have migrated their source code repositories and issue tracking systems into GitHub without a proper reassignment of bug report identifiers.
Therefore, for these projects, bug IDs referred to in the commit logs are ambiguous (for some bugs this may match with the GitHub issue tracking numbering, while in others, it refers to the original issue tracker). To avoid introducing noise in our validation data, we simply drop these projects.
For the remaining projects (Lang and Math), we leverage the bug linking strategies implemented in the Jira issue tracking software. We use a similar approach to Fischer et al.~\cite{fischer2003populating} and Thomas et al.~\cite{thomas2013impact} to link to commits to corresponding bug reports.
Concretely, we crawled the bug reports related to each project and assessed the links with a two-step search strategy: (i) we check commit logs to identify bug report IDs and associate the corresponding changes as bug fix changes; then (ii) we check for bug reports that are indeed considered as such (i.e., tagged as "BUG") and are further marked as resolved (i.e., with tags "RESOLVED" or "FIXED"), and completed (i.e., with status "CLOSED").

Eventually, our evaluation dataset includes {\bf 156 faults} (i.e., Defects4J bugs). Actually, for the considered projects, Defects4J enumerates 171 bugs associated with {\bf 162 bug reports}: 15 bugs are indeed left out because either (1) the corresponding bug reports are not in the desired status in the bug tracking system, which may lead to noisy data, or (2) there is ambiguity in the buggy program version (e.g., some fixed files appear to be missing in the repository at the time of bug reporting).

\subsubsection{Test suite reorganization} We ensure that the benchmark separates past test cases (i.e., regression test cases) from future test cases (i.e., test cases that encode functional requirements specified after the bug is reported). This timeline split is necessary to simulate the snapshot of the repository at the time the bug is reported.
As highlighted in Section~\ref{sec:motivation}, for over 90\% cases of bugs in the Defects4J benchmark, the test cases relevant to the defective behavior was actually provided along the bug fixing patches.
We have thus manually split the commits to identify test cases that should be considered as future test cases for each bug report.

\subsection{Implementation Choices}
\label{subsec:implementation}
During implementation, we have made the following parameter choices in the \toolname workflow:
\begin{itemize}[leftmargin=*]
	\item IR fault localization considers the top~50 (i.e., $k=50$ in Algorithm~\ref{alg:file2stmt}) suspicious files for each bug report, in order to search for buggy code locations.
	\item For patch recommendation experiments, we limit the search space to the top~20 suspected buggy statements yielded by the fine-grained IR-based fault localization.
	\item For comparison experiments, we implement spectrum-based fault localization using the GZoltar testing framework with the Ochiai ranking strategy. Unless otherwise indicated, GZoltar version 0.1.1 is used (as it is widely adopted in the literature, by Astor~\cite{martinez2016astor},
ACS~\cite{xiong2017precise}, ssFix~\cite{xin2017leveraging}
and CapGen~\cite{wen2018context} among others).
\end{itemize}

\subsection{Research Questions}

The assessment objective is to assess the {\bf feasibility of automating the generation of patches for user-reported bugs}, while investigating the foreseen bottlenecks as well as the research directions that the community must embrace to realize this long-standing endeavor. To that end, we focus on the following research questions associated with the different steps in the \toolname workflow.

\begin{itemize}[leftmargin=*]
	\item RQ1 [Fault localization] : {\em To what extent does IR-based fault localization provide reliable results for an APR scenario?} In particular, we investigate the performance differences when comparing our fine-grained IRFL implementation against the classical spectrum-based localization.
	\item RQ2 [Overfitting] : {\em To what extent does IR-based fault localization point to locations that are less subject to overfitting?} In particular, we study the impact on the {\em overfitting} problem that incomplete test suites generally carry.
 	\item RQ3 [Patch ordering] : {\em What is the effectiveness of \toolname's patch ordering strategy?} In particular, we investigate the overall workflow of \toolname, by re-simulating the real-world cases of software maintenance cycle when a bug is reported: future test cases are not available for patch validation.
 	\end{itemize}

\section{Assessment Results}
\label{sec:results}
In this section, we present the results of the investigations for the previously-enumerated research questions.

\subsection{RQ1: [Fault Localization]}
Fault localization being the first step in program repair, we evaluate the performance of the IR-based fault localization developed within \toolname. As recently thoroughly studied by Liu et al.~\cite{liu2019you}, an APR tool should not be expected to fix a bug that current fault localization systems fail to localize.
Nevertheless, with \toolname, we must demonstrate that our fine-grained IRFL offers comparable performance with SBFL tools used in the APR literature.

Table~\ref{tab:flresults} provides performance measurements on the localization of bugs.
SBFL is performed based on two different versions of the GZoltar testing framework, but always based on the Ochiai ranking metric.
Finally, because fault localization tools output a ranked list of suspicious statements, results are provided in terms of whether the correct location is placed under the top-k suspected statements. In this work, following the practice in the literature~\cite{thung2012faults,liu2019you}, we consider that a bug is localized if any buggy statement is localized.

\begin{table}[!h]
	\centering
	\scriptsize
	\caption{Fault localization results: {IRFL (IR-based) vs. SBFL  (Spectrum-based) on Defects4J (Math and Lang) bugs.}}
	\begin{threeparttable}
		\begin{tabular}{L{8mm}|L{8mm}|cccccC{9mm}}
			\toprule
			\multicolumn{2}{l|}{(171 bugs)}& Top-1 & Top-10  & Top-50  &Top-100  &Top-200  & All\\
			\hline
			\multicolumn{2}{l|}{{\bf IRFL}}  & 25 & 72 & 102 & 117 & 121 & 139 \\\hline
			\multirow{2}{*}{{\bf SBFL}} & {\bf{GZ$_{v1}$}} & 26 & 75 & 106 & 110 & 114 & 120\\
			& {\bf{GZ$_{v2}$}} & 23 & 79 & 119 & 135 & 150 & 156 \\
			\bottomrule
		\end{tabular}
		{$^\dagger$ GZ$_{v1}$ and GZ$_{v2}$ refer to GZoltar 0.1.1 and 1.6.0 respectively, which are widely used in APR systems for Java programs.}
	\end{threeparttable}
	\label{tab:flresults}
\end{table}

\begin{figure}[!ht]
\resizebox{0.4\linewidth}{!}{%
  \includegraphics[width=\linewidth]{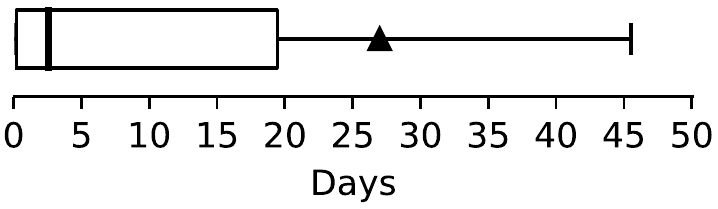}
	}
\caption{Distribution of elapsed time (in days) between bug report submission and test case attachment.}
\label{fig:gain}
\end{figure}

Overall, the results show that our IRFL implementation is strictly comparable to the common implementation of spectrum-based fault localization when applied on the Defects4J bug dataset.
Note that the comparison is conducted for 171 bugs of Math and Lang, given that these are the projects for which the bug linking can be reliably performed for applying the IRFL.
Although performance results are similar, we remind the reader that SBFL is applied by considering future test cases.
To highlight a practical interest of IRFL, we compute for each bug localizable in the top-10, the elapsed time between the bug report date and the date the relevant test case is submitted for this bug.
Based on the distribution shown in Figure~\ref{fig:gain}, on mean average, IRFL could reduce this time by 26 days.

Finally, to stress the importance of future test cases for spectrum-based fault localization, we consider all Defects4J bugs and compute localization performance with and without future test cases.

 Results listed in
Table~\ref{tab:gzoltar} confirms that in most bug cases, the localization is impossible: Only 10 bugs (out of 395) can be localized among the top-10 suspicious statements of SBFL at the time the bug is reported. In comparison, our IRFL locates 72 bugs under the same conditions of having no relevant test cases to trigger the bugs.

\begin{table}[!ht]
	\centering
	\scriptsize
	\caption{Fault localization performance.}
	\begin{threeparttable}
		\begin{tabular}{l|c|c|c|c|c|c}
			\toprule
			{\bf GZoltar + Ochiai} (395 bugs)& Top-1 & Top-10 &  Top-50 & Top-100 & Top-200 & All\\
			\hline
			without future tests        & 5 & 10 & 17 & 17 & 19 & 20\\
			with future tests & 45& 140& 198& 214& 239 & 263\\
			\bottomrule
		\end{tabular}
	\end{threeparttable}
	\label{tab:gzoltar}
\end{table}

\find{Fine-grained IR-based fault localization in \toolname is as accurate as Spectrum-based fault localization in localizing Defects4J bugs. Additionally, it does not have the constraint of requiring test cases that may not be available when the bug is reported.}

\subsection{RQ2: [Overfitting]}
Patch generation attempts to mutate suspected buggy code with suitable fix patterns.
Aside from having adequate patterns or not (which is out of the scope of our study), a common challenge of APR lies in the effective selection of buggy statements. In typical test-based APR, test cases drive the selection of these statements. The incompleteness of test suites is however currently suspected to often lead to overfitting of generated patches~\cite{yang2017better}.

We perform patch generation experiments to investigate the impact of localization bias. We compare our IRFL implementation against commonly-used SBFL implementations in the literature of test-based APR. We recall that the patch validation step in these experiments makes no assumptions about future test cases (i.e., all test cases are leveraged as in classical APR pipeline). For each bug, depending on the rank of the buggy statements in the suspicious statements yielded the fault localization system (either IRFL or SBFL), the patch generation can produce more or less relevant patches.
Table~\ref{tab:repairPerf} details the repair performance in relation to the position of buggy statements in the output of fault localization. Results are provided in terms of numbers of {\em plausible} and {\em correct}~\cite{qi2015analysis} patches that can be found by considering top-$k$ statements returned by the fault localizer.

\begin{table}[!ht]
	\centering
	\scriptsize
	\caption{IRFL vs. SBFL impacts on the number of generated correct/plausible patches for Defects4J bugs.}
	\begin{threeparttable}
		\begin{tabular}{L{18mm}|C{15mm}C{15mm}|C{15mm}}
			\toprule
			 & {\bf Lang} & {\bf Math} & {\bf Total}\\
			\hline
			\rowcolor{lightgray}IRFL Top-1 & 1/4 & 3/4 & 4/8 \\
			SBFL  Top-1 & 1/4 & {\bf6/8} & 7/12\\
			\hline
			\hline
			\rowcolor{lightgray}IRFL Top-5 & {\bf3}/6 & 7/14 & 10/20 \\
			SBFL Top-5 & 2/{\bf7} & {\bf11/17} & 13/24\\
			\hline
			\hline
			\rowcolor{lightgray}IRFL Top-10 & {\bf4/9} & 9/17 & 13/26 \\
			SBFL Top-10 & 4/11 & {\bf16/27} & 20/38\\
			\hline
			\hline
			\rowcolor{lightgray}IRFL Top-20 & {\bf7/12} & 9/18 & 16/30 \\
			SBFL Top-20 & 4/11 & {\bf18/30} & 22/41\\
			\hline
			\hline
			\rowcolor{lightgray}IRFL Top-50 & {\bf7/15} & 10/22 & 17/37 \\
			SBFL Top-50 & 4/13 & {\bf19/34} & 23/47\\
			\hline
			\hline
			\rowcolor{lightgray}IRFL Top-100 & {\bf8/18} & 10/23 & 18/41 \\
			SBFL Top-100 & 5/14 & {\bf19/36} & 24/50\\
			\hline
			\hline
			\rowcolor{lightgray}IRFL All & {\bf11/19} & 10/25 & 21/44 \\
			SBFL All & 5/14 & {\bf19/36} & 24/50\\
			\bottomrule
		\end{tabular}
		{$^\ast$ We indicate x/y numbers of patches: x is the number of bugs for which a {\em correct} patch is generated; y is the number of bugs for which a {\em plausible} patch is generated.}
	\end{threeparttable}
	\label{tab:repairPerf}
\end{table}

Overall, we find that IRFL and SBFL localization information lead to similar repair performance in terms of the number of fixed bugs plausibly/correctly. Actually IRFL-supported APR outperforms SBFL-supported APR on the Lang project bugs and vice-versa for Math project bugs: overall, 6 bugs that are fixed using IRFL output, cannot be fixed using SBFL output (although assuming the availability of the bug triggering test cases to run the SBFL tool).

We investigate the cases of plausible patches in both localization scenarios to characterize the reasons why these patches appear to only be overfitting the test suites. Table~\ref{tab:dissectionOfPlausible} details the overfitting reasons for the two scenarios.
\begin{table}[!h]
	\centering
	\scriptsize
	\caption{Dissection of reasons why patches are plausible$^*$ but not correct.}
	\begin{threeparttable}
		\begin{tabular}{L{8mm}|c|c|c}
			\toprule
			& \makecell[c]{Localization Error} &  \makecell[c]{Pattern Prioritization} & \makecell[c]{Lack of Fix ingredients}  \\
			\hline
			w/ IRFL & 6 & 1  & 16 \\
			w/ SBFL & 15 & 1  & 10 \\
			\bottomrule
		\end{tabular}
{$^*$A plausible patch passes all test cases, but may not be semantically equivalent to developer patch (i.e., correct). We consider a plausible patch to be overfitted to the test suite}
	\end{threeparttable}
	\label{tab:dissectionOfPlausible}
\end{table}

\begin{enumerate}[leftmargin=*]
\item Among the $23 (= 44-21)$ plausible patches that are generated based on IRFL identified code locations and that are not found to be correct, 6 are found to be caused by fault localization errors: these bugs are plausibly fixed by mutating irrelevantly-suspicious statements that are placed before the actual buggy statements in the fault localization output list.
This phenomenon has been recently investigated in the literature as the problem of fault localization bias~\cite{liu2019you}.
Nevertheless, we note that patches generated based on SBFL identified code locations suffer more of fault localization bias: 15 of the 26 (= 50$-$24) plausible patches are concerned by this issue.

\item Pattern prioritization failures may also lead to plausible patches: while a correct patch could have been generated using a specific pattern at a lower node in the AST, another pattern (leading to an only plausible patch) was first found to be matching the statement during the iterative search of matching nodes (cf. Section~\ref{subsec:generation}).

\item Finally, we note that both configurations yield plausible patches due to the lack of suitable patterns or due to a failed search for the adequate donor code (i.e., fix ingredient~\cite{liu2018lsrepair}).
\end{enumerate}

\find{Experiments with the Defects4J dataset suggest that code locations provided by IR-based fault localization lead less to overfitted patches than the code locations suggested by Spectrum-based fault localization: cf. "Localization error" column in Table~\ref{tab:dissectionOfPlausible}.}

\subsection{RQ3: [Patch Ordering]}
While the previous experiment focused on patch generation, our final experiment assesses the complete pipeline of \toolname as it was imagined for meeting the constraints that developers can face in practice: future test cases, i.e., those which encode the functionality requirements that are not met by the buggy programs, may not be available at the time the bug is reported.
We thus discard the future test cases of the Defects4J dataset and generate patches that must be recommended to developers. The evaluation protocol thus consists in assessing to what extent correct/plausible patches are  placed in the top of the recommendation list.


\subsubsection{Overall performance}
Table~\ref{tab:suggestion} details the performance of the patch recommendation by \toolname: we present the number of bugs for which a correct/plausible patch is generated and presented among the top-$k$ of the list of recommended patches.
In the absence of future test cases to drive the patch validation process, we use heuristics (cf. Section~\ref{subsec:implementation}) to re-prioritize the patch candidates towards ensuring that patches which are recommended first will eventually be correct (or at least plausible when relevant test cases are implemented). We present results both for the case where we do not re-prioritize and the case where we re-prioritize.

Recall that, given that the re-organized benchmark separately includes the future test cases, we can leverage them to systematize the assessment of patch plausibility. The {\em correctness} (also referred to as {\em correctness}~\cite{qi2015analysis}) of patches, however, is still decided manually by comparing against the actual bug fix provided by developers and available in the benchmark. Overall, we note that \toolname performance is promising as it manages, for {\bf 13 bugs}, to present a plausible patch among its top-5 recommended patches per bug. Among those plausible patches, 8 are eventually found to be correct.

\begin{table}[!h]
	\centering
	\scriptsize
	\caption{Overall performance of \toolname for patch recommendation on the Defects4J benchmark.}
	\label{tab:suggestion}
	\begin{threeparttable}
		\begin{tabular}{l|ccccc}
			\toprule
			Recommendation rank & Top-1 & Top-5 & Top-10 & Top-20 & All\\
			\hline
			{\bf without} patch re-prioritization & 3/3 & 4/5  & 6/10 & 6/10 & 13/27\\
			{\bf with} patch re-prioritization    & 3/4 & 8/13 & 9/14 & 10/15& 13/27\\
			\bottomrule
		\end{tabular}
		{$^\ast$ x/y: x is the number of bugs for which a {\em correct} patch is generated; y is the number of bugs for which a {\em plausible} patch is generated.}
	\end{threeparttable}
\end{table}

\subsubsection{Comparison with the state-of-the-art test-based APR systems}
To objectively position the performance of \toolname (which does not require future test cases to localize bugs, generate patches and present a sorted recommendation list of patches), we count the number of bugs for which \toolname can propose a correct/plausible patch.
We consider three scenarios with \toolname:
\begin{enumerate}[leftmargin=*]
	\item {[$\toolname_{top5}$]} - developers will be provided with only top 5 recommended patches which have been validated only with regression tests: in this case, \toolname outperforms about half of the state-of-the-art in terms of numbers bugs fixed with both plausible or correct patches.
	\item {[$\toolname_{all}$]} - developers are presented with all (i.e., not only top-5) generated patches validated with regression tests: in this case, only four (out of sixteen) state-of-the-art APR techniques outperform \toolname.
	\item {[$\toolname_{opt}$]} - developers are presented with all generated patches which have been validated with augmented test suites (i.e., optimistically with future test cases): with this configuration, \toolname outperforms all state-of-the-art, except SimFix~\cite{jiang2018shaping} which uses sophisticated techniques to improve the fault localization accuracy and search for fix ingredients. It should be noted that in this case, our prioritization strategy is not applied to the generated patches. $\toolname_{opt}$ represents the reference performance for our experiment which assesses the prioritization.
\end{enumerate}
Table~\ref{tab:sota} provides the comparison matrix. Information on state-of-the-art results are excerpted from their respective publications.
\begin{table}[!h]
	\centering
	\scriptsize
	\caption{\toolname vs state-of-the-art APR tools.}
	\label{tab:sota}
	\begin{threeparttable}
		\begin{tabular}{L{20mm}|C{16mm}C{16mm}C{16mm}}
			 \toprule
			 APR tool & {\bf Lang$^\ast$} & {\bf Math$^\ast$} & {\bf Total$^\ast$}\\
			 \hline
			 jGenProg~\cite{martinez2016astor} & 0/0 & 5/18 & 5/18 \\
			 jKali~\cite{martinez2016astor}    & 0/0 & 1/14 & 1/14 \\
		   jMutRepair~\cite{martinez2016astor} & 0/1 & 2/11 & 2/12 \\
		     HDRepair~\cite{le2016history}     & 2/6 & 4/7  & 6/13 \\
		     Nopol~\cite{xuan2017nopol}        & 3/7 & 1/21 & 4/28 \\
		     ACS~\cite{xiong2017precise}       & 3/4 & 12/16& 15/20\\
		     ELIXIR~\cite{saha2017elixir}      & 8/12& 12/19& 20/31\\
		     JAID~\cite{chen2017contract}      & 1/8 & 1/8  & 2/16 \\
		     ssFix~\cite{xin2017leveraging}    & 5/12& 10/26& 15/38\\
		     CapGen~\cite{wen2018context}      & 5/5 & 12/16& 17/21\\
		     SketchFix~\cite{hua2018towards}   & 3/4 & 7/8  & 10/12\\
		   FixMiner~\cite{koyuncu2018fixminer} & 2/3 & 12/14& 14/17\\
		     LSRepair~\cite{liu2018lsrepair}   & 8/14& 7/14 & 15/28\\
		     SimFix~\cite{jiang2018shaping}    & 9/13& {\bf14/26}& {\bf23}/39\\
		     kPAR~\cite{liu2019you}            & 1/8 & 7/18 & 8/26\\
		     AVATAR~\cite{liu2019avatar}       & 5/11& 6/13 & 11/24\\
		     \hline
		     $\toolname_{opt}$ & {\bf11/19} & 10/25 & 21/{\bf44}\\ \hline
		     $\toolname_{all}$ & 6/11 & 7/16 & 13/27\\
		     $\toolname_{top5}$ & 3/7 & 5/6 & 8/13\\
			 \bottomrule
		\end{tabular}
		{$^\ast$ $x/y$: x is the number of bugs for which a {\em correct} patch is generated; y is the number of bugs for which  a {\em plausible} patch is generated.\\
		$\toolname_{opt}$: the version of \toolname where available test cases are relevant to the bugs.\\
		$\toolname_{all}$: all recommended patches are considered.\\
		$\toolname_{top5}$: only top 5 recommended patches are considered.}
	\end{threeparttable}
\end{table}

\find{\toolname offers a reasonable performance in patch recommendation when we consider the number of Defects4J bugs that are successfully patched among the top-5 (in a scenario where we assume not having relevant test cases to validate the patch candidates). Performance results are even comparable to many state-of-the-art test-based APR tools in the literature.}

\subsubsection{Properties of \toolname's patches}
In Table~\ref{tab:complexity}, we characterize the correct and plausible patches recommended by $\toolname_{top5}$. Overall, update and insert changes have been successful; most patches affect a single statement, and impact precisely an expression entity within a statement.

\begin{table}[!h]
	\centering
	\scriptsize
	\caption{Change properties of \toolname's correct patches.}
	\label{tab:complexity}
	\begin{threeparttable}
		\begin{tabular}{lc||lc||lc}
			\toprule
			Change action & \#bugs$^\ast$ & Impacted statement(s)	& \#bugs$^\ast$ & Granularity & \#bugs$^\ast$\\
			\hline
			Update & 5/7 & Single-statement   & 8/12 & Statement  & 1/2\\
		 	Insert & 3/5 & Multiple-statement &  0/1 & Expression & 7/11\\
			Delete & 0/1\\
			\bottomrule
		\end{tabular}
		{$^\ast$ x/y $\longrightarrow$  for x bugs the patches are correct, while for y bugs they are plausible.}
	\end{threeparttable}
\end{table}

\subsubsection{Diversity of \toolname's fixed bugs} Finally, in Table~\ref{tab:dissectionBugReports} we dissect the nature of the bugs for which $\toolname_{top5}$ is able to recommend a correct or a plausible patch. Priority information about the bug report is collected from the issue tracking systems, while the root cause is inferred by analyzing the bug reports and fixes.

\begin{table}[!h]
	\centering
	\scriptsize
	\caption{Dissection of bugs successfully fixed by \toolname.}
	\resizebox{1\linewidth}{!}{
	\begin{threeparttable}
		\begin{tabularx}{\linewidth}{c|c|c|X|c}
			\toprule
			\makecell[l]{Patch \\ Type }& \makecell[l]{Defect4J \\ Bug ID} &Issue ID&  \makecell[l]{Root Cause} & \makecell[l]{Priority}  \\
			\hline
				G & L-6 	& LANG-857	& {String index out of bounds exception}& Minor \\
				G & L-24	& LANG-664  & {Wrong behavior due missing condition}   & Major\\
				G & L-57	& LANG-304	& {Null pointer exception}					& Major \\
				G & M-15	& MATH-904	& {Double precision floating point format error}	& Major \\
				G & M-34	& MATH-779	& {Missing "read only access" to internal list}		& Major \\
				G & M-35	& MATH-776	& {Range check}								& Major \\
				G & M-57	& MATH-546	& {Wrong variable type truncates double value}	& Minor \\
				G & M-75	& MATH-329	& {Method signature mismatch} &	Minor \\
				P & L-13    & LANG-788	& {Serialization error in primitive types}	& Major \\
				P & L-21 	& LANG-677 	& {Wrong Date Format in comparison} & Major \\
				P & L-45    & LANG-419	& {Range check}	& Minor \\
				P & L-58	& LANG-300	& {Number formatting error} & 	Major \\
				P & M-2		& MATH-1021	& {Integer overflow}	& Major \\
			\bottomrule
		\end{tabularx}
		{``G'' denotes correct patch and ``P'' means plausible patch.}
	\end{threeparttable}
	}
	\label{tab:dissectionBugReports}
\end{table}

Overall, we note that 9 out of the 13 bugs have been marked as Major issues. 12 different bug types (i.e., root causes) are addressed. In contrast, R2Fix~\cite{liu2013r2fix} only focused on 3 simple bug types.

\section{Discussion}
This study presents the conclusions of our investigation into the feasibility of generating patches automatically from bug reports. We set strong constraints on the absence of test cases, which are used in test-based APR to approximate {\em what the program is actually supposed to do} and {\em when the repair is completed}~\cite{westley2009automatically}.
Our experiments on the widely-used Defects4J bugs eventually show that \underline{\em patch generation without bug-triggering test cases} is promising.

Manually looking at the details of failures and success in generating patches with \toolname, several insights can be drawn:

\noindent
{\bf Test cases can be buggy:}
During manual analysis of results, we noted that \toolname actually fails to generate correct patches for three bugs (namely, Math-5, Math-59 and Math-65) because even the test cases were buggy. Figure~\ref{fig:math5} illustrates the example of bug Math-5 where its patch also updated the relevant test case. This example supports our endeavor, given that users would find and report bugs for which the appropriate test cases were never properly written.

\begin{figure}[!h]
    \centering\vspace{1mm}
\begin{lstlisting}[language=diff,linewidth={\linewidth},frame=tb,basicstyle=\footnotesize\ttfamily]
// (@{\bf\ttfamily Patched Source Code: }@)
--- a/src/main/java/org/apache/commons/math3/complex/Complex.java
+++ b/src/main/java/org/apache/commons/math3/complex/Complex.java
@@ -304,3 +304,3 @@ public class Complex
         if (real == 0.0 && imaginary == 0.0) {
(@{\textcolor{red}{-}@)            return (@{\textcolor{red}{NaN}@);
(@{\textcolor{codegreen}{+}@)            return (@{\textcolor{codegreen}{INF}@);
         }

// (@{\bf\ttfamily Patched Test Case: }@)
(@{--- a/src/test/java/org/apache/commons/math3/complex/ComplexTest.java}@)
(@{+++ b/src/test/java/org/apache/commons/math3/complex/ComplexTest.java}@)
@@ -333,3 +333,3 @@ public class ComplexTest {
     public void testReciprocalZero() {
(@{\textcolor{red}{-}@)       Assert.assertEquals(Complex.ZERO.reciprocal(), Complex.(@{\textcolor{red}{NaN}@));
(@{\textcolor{codegreen}{+}@)       Assert.assertEquals(Complex.ZERO.reciprocal(), Complex.(@{\textcolor{codegreen}{INF}@));
     }
\end{lstlisting}
    \caption{Patched source code and test case of fixing Math-5.}
    \label{fig:math5}
\end{figure}

\noindent
{\bf Bug reports deserve more interest:}
With \toolname, we have shown that bug reports could be handled automatically for a variety of bugs.
This is an opportunity for issue trackers to add a recommendation layer to the bug triaging process
by integrating patch generation techniques.
There are, however, several directions to further investigation, among which:
(1) help users write proper bug reports;
and (2) re-investigate 
IRFL techniques
at a finer-grained level that is suitable for APR.

\noindent
{\bf Prioritization techniques must be investigated:} In the absence of complete test suites for validating every single patch candidate, a recommendation system must ensure that patches presented first to the developers are the most likely to be plausible and even correct. There are thus two directions of research that are promising: (1) ensure that fix patterns are properly prioritized to generate good patches and be able to early-stop for not exploding the search space; and (2) ensure that candidate patches are effectively re-prioritized. These investigations must start with a thorough dissection of plausible patches for a deep understanding of plausibility factors.

\noindent
{\bf More sophisticated approaches to triaging and selecting fix ingredients are necessary:} In its current form, \toolname implements a na{\"i}ve approach to patch generation, ensuring that the performance is tractable. However, the literature already includes novel APR techniques that implement strategies for selecting donor code and filters patterns. Integrating such techniques into \toolname may lead to performance improvement.

\noindent
{\bf More comprehensive benchmarks are needed:} Due to bug linking challenges, our experiments were only performed on half of the Defects4J benchmark. To drive strong research in patch generation for user-reported bugs, the community must build larger and reliable benchmarks, potentially even linking several artifacts of continuous integration (i.e, build logs, past execution traces, etc.). In the future, we plan to investigate the dataset of Bugs.jar~\cite{saha2018bugs}.

\noindent
{\bf Automatic test generation techniques could be used as a supplement:} Our study tries to cope radically with the incompleteness of test suites. In the future, however, we could investigate the use of automatic test generation techniques to supplement the regression test cases during patch validation.

\section{Threats to Validity}
\begin{figure}[!ht]
	\includegraphics[width=0.8\linewidth]{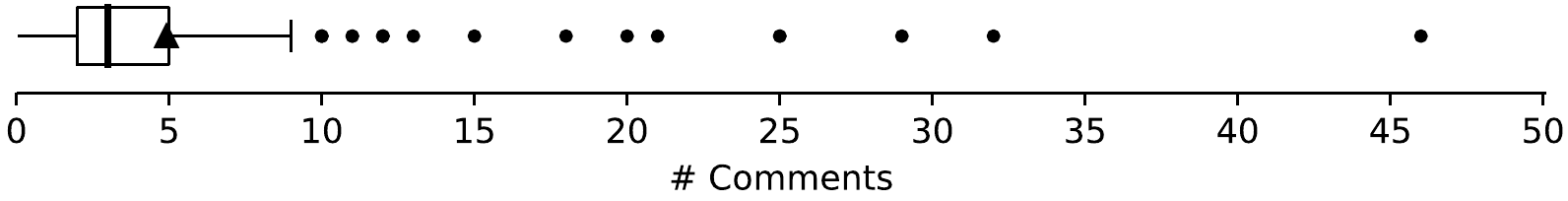}
	\caption{Distribution of \# of comments per bug report.}
	\label{fig:comment}
\end{figure}

\noindent
{\bf Threats to external validity:} The bug reports used in this study may be of low quality (i.e., wrong links for corresponding bugs). We reduced this threat by focusing only on bugs from the Lang and Math projects, which kept a single issue tracking system. We also manually verified the links between the bug reports and the Defects4J bugs. Table~\ref{tab:brQuality} characterizes the bug reports of our dataset following the criteria enumerated by Zimmermann et al.~\cite{zimmermann2010makes} in their study of ``what makes a good bug report''. Notably, as illustrated by the distribution of comments in Figure~\ref{fig:comment}, we note that the bug reports have been actively discussed before being resolved. This suggests that they are not trivial cases~(cf. \cite{hooimeijer2007modeling} on measuring bug report significance).

\begin{table}[!h]
	\centering
	\scriptsize
	\caption{Dissection of bug reports related to Defects4J bugs.}
	\resizebox{\linewidth}{!}{
		\begin{threeparttable}
			\begin{tabular}{lc||ccccccc}
				\toprule
				Proj. & \makecell[c]{ Unique\\ Bug Reports }  &  \makecell[c]{w/ Patch \\ Attached }  & \makecell[c]{Average\\Comments } & \makecell[c]{ w/ Stack \\Traces }  &  \makecell[c]{ w/ \\ Hints }    & \makecell[c]{ w/ Code\\Blocks }  \\
				\midrule
				Lang &62&            11 &         4.53 &               4 &               62 &            31 \\
				Math &100&           23 &         5.15 &               5 &               92 &            51 \\
				\bottomrule
			\end{tabular}
			{Code-related terms such as package/class names found in the summary and description, in addition to stack traces and code blocks, as separate features referred to as hints.}
		\end{threeparttable}
	}
	\label{tab:brQuality}
\end{table}

Another threat to external validity relates to the diversity of the fix patterns used in this study.
\toolname currently may not implement a reasonable number of relevant fix patterns. We minimize this threat by surveying the literature and considering patterns from several pattern-based APR.

\noindent
{\bf Threats to internal validity:}
Our implementation of fine-grained IRFL carries some threats: during the search of buggy statements,
we considered top-50 suspicious buggy files from the file-level IRFL tool, to limit the search space. Different threshold values may lead to different results. We also considered only 5 statement types as more bug-prone. This second threat is minimized by the empirical evidence provided by Liu et al.~\cite{liu2018closer}.

Additionally, another internal threat is in our patch generation steps: \toolname only searches for donor code from the local code files, which contain the buggy statement. The adequate fix ingredient may however be located elsewhere.

\noindent
{\bf Threats to construct validity:}
In this study, we assumed that patch construction and test case creation are two separated tasks for developers. This may not be the case in practice. The threat is however mitigated given that, in any case, we have shown that the test cases are often unavailable when the bug is reported.

\section{Related Work}


\noindent
{\bf Fault Localization.}
As stated in a recent study~\cite{liu2019you},
fault localization is a critical task affecting the effectiveness of automated program repair.
Several techniques have been proposed~\cite{parnin2011automated,wang2015evaluating,wong2016survey}
and they use different information such as spectrum~\cite{abreu2009spectrum}, text~\cite{wen2016locus}, slice~\cite{mao2014slice},
and statistics~\cite{liblit2005scalable}.
The first two types of techniques are widely studies in the community.
SBFL techniques~\cite{abreu2009practical,jones2005empirical} are widely adopted in APR pipelines
since they identify bug positions at a fine-grained level (i.e., statements).
However, they have limitations on localizing buggy locations since it highly relies on the test suite~\cite{liu2019you}.
Information retrieval based fault localization (IRFL)~\cite{lee2018bench4bl} leverages
textual information in a bug report. It is mainly used to help developers narrow down suspected buggy  files in the absence of relevant test cases. For the purpose of our study, we have proposed an algorithm for further localizing the faulty code entities at the statement level.

\noindent
{\bf Patch Generation.}
Patch generation is another key process of APR pipeline, which
is, in other words, a task searching for another shape of a program (i.e., a patch)
in the space of all possible programs~\cite{le2012systematic,long2016analysis}.
To improve repair performance, many APR systems have been explored to address the search space problem by using different information and approaches:
stochastic mutation~\cite{westley2009automatically,le2012genprog},
synthesis~\cite{long2015staged,xuan2017nopol,xiong2017precise},
pattern~\cite{kim2013automatic,le2016history,saha2017elixir,long2017automatic, durieux2017dynamic,le2017s3,liu2018mining,hua2018towards,jiang2018shaping,liu2019avatar},
contract~\cite{wei2010automated,chen2017contract},
symbolic execution~\cite{nguyen2013semfix},
learning~\cite{long2016automatic,gupta2017deepfix,rolim2017learning,
soto2018using,bhatia2018neuro,white2019sorting},
and donor code searching~\cite{mechtaev2015directfix, ke2015repairing}.
In this paper, patch generation is implemented with fix patterns presented in the literature
since it may make the generated patches more robust~\cite{schulte2014software}.

\noindent
{\bf Patch Validation.}
The ultimate goal of APR systems is to automatically generate a \emph{correct} patch
that can actually resolve the program defects rather than satisfying minimal
functional constraints.
At the beginning, patch correctness is evaluated by passing all test
cases~\cite{westley2009automatically,kim2013automatic,le2016history}.
However, these patches could be overfitting~\cite{qi2015analysis,le2018overfitting}
and even worse than the bug~\cite{smith2015cure}.
Since then, APR systems are evaluated with the precision of generating correct
patches~\cite{xiong2017precise,wen2018context,jiang2018shaping,liu2019avatar}.
Recently, researchers explore automated frameworks that can identify
patch correctness for APR systems automatically~\cite{xiong2018identifying, le2019reliability}.
In this paper, our approach validates generated patches with regression test suites
since fail-inducing test cases are readily available for most of bugs as described in Section~\ref{sec:motivation}.

\section{Conclusion}
In this study, we have investigated the feasibility of automating patch generation
from bug reports. To that end, we implemented \toolname, an APR pipeline variant
adapted to the constraints of test cases unavailability
when users report bugs. The proposed system revisits the fundamental steps,
notably fault localization, patch generation and patch validation,
which are all tightly-dependent to the {\em positive test cases}~\cite{westley2009automatically}
in a test-based APR system.

Without making any assumptions on the availability of test cases, we demonstrate, after re-organizing the Defects4J benchmark, that \toolname can generate and recommend priority correct (and more plausible) patches for a diverse set of user-reported bugs. The repair performance of \toolname is even found to be comparable to that of the majority of test-based APR systems on the Defects4J dataset.

We open source \toolname's code and release all data of this study to facilitate replication and encourage further research in this direction which is promising for practical adoption in the software development community:
\begin{center}
		\url{https://github.com/SerVal-DTF/iFixR}
\end{center}

\begin{acks}
This work is supported by the Fonds National de la Recherche (FNR), Luxembourg, through RECOMMEND 15/IS/10449467 and FIXPATTERN C15/IS/9964569.
\end{acks}

\balance
\bibliographystyle{ACM-Reference-Format}
\bibliography{bib/references}


\begin{thebibliography}{100}


\ifx \showCODEN    \undefined \def \showCODEN     #1{\unskip}     \fi
\ifx \showDOI      \undefined \def \showDOI       #1{#1}\fi
\ifx \showISBNx    \undefined \def \showISBNx     #1{\unskip}     \fi
\ifx \showISBNxiii \undefined \def \showISBNxiii  #1{\unskip}     \fi
\ifx \showISSN     \undefined \def \showISSN      #1{\unskip}     \fi
\ifx \showLCCN     \undefined \def \showLCCN      #1{\unskip}     \fi
\ifx \shownote     \undefined \def \shownote      #1{#1}          \fi
\ifx \showarticletitle \undefined \def \showarticletitle #1{#1}   \fi
\ifx \showURL      \undefined \def \showURL       {\relax}        \fi
\providecommand\bibfield[2]{#2}
\providecommand\bibinfo[2]{#2}
\providecommand\natexlab[1]{#1}
\providecommand\showeprint[2][]{arXiv:#2}

\bibitem[\protect\citeauthoryear{??}{dAN}{2019}]%
        {dANDc}
 \bibinfo{year}{2019}\natexlab{}.
\newblock \bibinfo{title}{D\&C}.
\newblock \bibinfo{howpublished}{\url{https://github.com/d-and-c/d-and-c}}.
\newblock


\bibitem[\protect\citeauthoryear{Abreu, Van~Gemund, and Zoeteweij}{Abreu
  et~al\mbox{.}}{2007}]%
        {abreu2007accuracy}
\bibfield{author}{\bibinfo{person}{Rui Abreu}, \bibinfo{person}{Arjan~JC
  Van~Gemund}, {and} \bibinfo{person}{Peter Zoeteweij}.}
  \bibinfo{year}{2007}\natexlab{}.
\newblock \showarticletitle{On the accuracy of spectrum-based fault
  localization}. In \bibinfo{booktitle}{\emph{Proceedings of
  TAICPART-MUTATION}}. IEEE, \bibinfo{pages}{89--98}.
\newblock


\bibitem[\protect\citeauthoryear{Abreu, Zoeteweij, Golsteijn, and
  Van~Gemund}{Abreu et~al\mbox{.}}{2009b}]%
        {abreu2009practical}
\bibfield{author}{\bibinfo{person}{Rui Abreu}, \bibinfo{person}{Peter
  Zoeteweij}, \bibinfo{person}{Rob Golsteijn}, {and} \bibinfo{person}{Arjan~JC
  Van~Gemund}.} \bibinfo{year}{2009}\natexlab{b}.
\newblock \showarticletitle{A practical evaluation of spectrum-based fault
  localization}.
\newblock \bibinfo{journal}{\emph{JSS}} \bibinfo{volume}{82},
  \bibinfo{number}{11} (\bibinfo{year}{2009}), \bibinfo{pages}{1780--1792}.
\newblock


\bibitem[\protect\citeauthoryear{Abreu, Zoeteweij, and Van~Gemund}{Abreu
  et~al\mbox{.}}{2009a}]%
        {abreu2009spectrum}
\bibfield{author}{\bibinfo{person}{Rui Abreu}, \bibinfo{person}{Peter
  Zoeteweij}, {and} \bibinfo{person}{Arjan~JC Van~Gemund}.}
  \bibinfo{year}{2009}\natexlab{a}.
\newblock \showarticletitle{Spectrum-based multiple fault localization}. In
  \bibinfo{booktitle}{\emph{Proceedings of the 24th ASE}}. IEEE,
  \bibinfo{pages}{88--99}.
\newblock


\bibitem[\protect\citeauthoryear{Anvik, Hiew, and Murphy}{Anvik
  et~al\mbox{.}}{2006}]%
        {anvik2006should}
\bibfield{author}{\bibinfo{person}{John Anvik}, \bibinfo{person}{Lyndon Hiew},
  {and} \bibinfo{person}{Gail~C Murphy}.} \bibinfo{year}{2006}\natexlab{}.
\newblock \showarticletitle{Who should fix this bug?}. In
  \bibinfo{booktitle}{\emph{Proceedings of the 28th ICSE}}. ACM,
  \bibinfo{pages}{361--370}.
\newblock


\bibitem[\protect\citeauthoryear{Beck}{Beck}{2003}]%
        {beck2003test}
\bibfield{author}{\bibinfo{person}{Kent Beck}.}
  \bibinfo{year}{2003}\natexlab{}.
\newblock \bibinfo{booktitle}{\emph{Test-driven development: by example}}.
\newblock \bibinfo{publisher}{Addison-Wesley Professional}.
\newblock


\bibitem[\protect\citeauthoryear{Beller, Gousios, Panichella, and
  Zaidman}{Beller et~al\mbox{.}}{2015}]%
        {beller2015and}
\bibfield{author}{\bibinfo{person}{Moritz Beller}, \bibinfo{person}{Georgios
  Gousios}, \bibinfo{person}{Annibale Panichella}, {and} \bibinfo{person}{Andy
  Zaidman}.} \bibinfo{year}{2015}\natexlab{}.
\newblock \showarticletitle{When, how, and why developers (do not) test in
  their IDEs}. In \bibinfo{booktitle}{\emph{Proceedings of the 10th FSE}}. ACM,
  \bibinfo{pages}{179--190}.
\newblock


\bibitem[\protect\citeauthoryear{Bhatia, Kohli, and Singh}{Bhatia
  et~al\mbox{.}}{2018}]%
        {bhatia2018neuro}
\bibfield{author}{\bibinfo{person}{Sahil Bhatia}, \bibinfo{person}{Pushmeet
  Kohli}, {and} \bibinfo{person}{Rishabh Singh}.}
  \bibinfo{year}{2018}\natexlab{}.
\newblock \showarticletitle{Neuro-symbolic program corrector for introductory
  programming assignments}. In \bibinfo{booktitle}{\emph{Proceedings of the
  40th ICSE}}. ACM, \bibinfo{pages}{60--70}.
\newblock


\bibitem[\protect\citeauthoryear{Bissyand{\'e}}{Bissyand{\'e}}{2015}]%
        {bissyande2015harvesting}
\bibfield{author}{\bibinfo{person}{Tegawend{\'e}~F Bissyand{\'e}}.}
  \bibinfo{year}{2015}\natexlab{}.
\newblock \showarticletitle{Harvesting Fix Hints in the History of Bugs}.
\newblock \bibinfo{journal}{\emph{arXiv preprint arXiv:1507.05742}}
  (\bibinfo{year}{2015}).
\newblock


\bibitem[\protect\citeauthoryear{B{\"o}hme, Soremekun, Chattopadhyay,
  Ugherughe, and Zeller}{B{\"o}hme et~al\mbox{.}}{2017}]%
        {bohme2017bug}
\bibfield{author}{\bibinfo{person}{Marcel B{\"o}hme},
  \bibinfo{person}{Ezekiel~O Soremekun}, \bibinfo{person}{Sudipta
  Chattopadhyay}, \bibinfo{person}{Emamurho Ugherughe}, {and}
  \bibinfo{person}{Andreas Zeller}.} \bibinfo{year}{2017}\natexlab{}.
\newblock \showarticletitle{Where is the bug and how is it fixed? an experiment
  with practitioners}. In \bibinfo{booktitle}{\emph{Proceedings of the 11th
  FSE}}. ACM, \bibinfo{pages}{117--128}.
\newblock


\bibitem[\protect\citeauthoryear{Campos, Riboira, Perez, and Abreu}{Campos
  et~al\mbox{.}}{2012}]%
        {campos2012gzoltar}
\bibfield{author}{\bibinfo{person}{Jos{\'e} Campos}, \bibinfo{person}{Andr{\'e}
  Riboira}, \bibinfo{person}{Alexandre Perez}, {and} \bibinfo{person}{Rui
  Abreu}.} \bibinfo{year}{2012}\natexlab{}.
\newblock \showarticletitle{Gzoltar: an eclipse plug-in for testing and
  debugging}. In \bibinfo{booktitle}{\emph{Proceedings of the 27th ASE}}.
  IEEE/ACM, \bibinfo{pages}{378--381}.
\newblock


\bibitem[\protect\citeauthoryear{Chen, Pei, and Furia}{Chen
  et~al\mbox{.}}{2017}]%
        {chen2017contract}
\bibfield{author}{\bibinfo{person}{Liushan Chen}, \bibinfo{person}{Yu Pei},
  {and} \bibinfo{person}{Carlo~A Furia}.} \bibinfo{year}{2017}\natexlab{}.
\newblock \showarticletitle{Contract-based program repair without the
  contracts}. In \bibinfo{booktitle}{\emph{Proceedings of the 32nd ASE}}.
  \bibinfo{publisher}{{IEEE}}, \bibinfo{pages}{637--647}.
\newblock


\bibitem[\protect\citeauthoryear{Coker and Hafiz}{Coker and Hafiz}{2013}]%
        {coker2013program}
\bibfield{author}{\bibinfo{person}{Zack Coker} {and} \bibinfo{person}{Munawar
  Hafiz}.} \bibinfo{year}{2013}\natexlab{}.
\newblock \showarticletitle{Program transformations to fix C integers}. In
  \bibinfo{booktitle}{\emph{Proceedings of the 35th ICSE}}.
  \bibinfo{publisher}{{IEEE}/{ACM}}, \bibinfo{pages}{792--801}.
\newblock


\bibitem[\protect\citeauthoryear{Durieux, Cornu, Seinturier, and
  Monperrus}{Durieux et~al\mbox{.}}{2017}]%
        {durieux2017dynamic}
\bibfield{author}{\bibinfo{person}{Thomas Durieux}, \bibinfo{person}{Benoit
  Cornu}, \bibinfo{person}{Lionel Seinturier}, {and} \bibinfo{person}{Martin
  Monperrus}.} \bibinfo{year}{2017}\natexlab{}.
\newblock \showarticletitle{Dynamic patch generation for null pointer
  exceptions using metaprogramming}. In \bibinfo{booktitle}{\emph{Proceedings
  of the 24th SANER}}. IEEE, \bibinfo{pages}{349--358}.
\newblock


\bibitem[\protect\citeauthoryear{Fischer, Pinzger, and Gall}{Fischer
  et~al\mbox{.}}{2003}]%
        {fischer2003populating}
\bibfield{author}{\bibinfo{person}{Michael Fischer}, \bibinfo{person}{Martin
  Pinzger}, {and} \bibinfo{person}{Harald Gall}.}
  \bibinfo{year}{2003}\natexlab{}.
\newblock \showarticletitle{Populating a release history database from version
  control and bug tracking systems}. In \bibinfo{booktitle}{\emph{Proceeding of
  the 19th ICSM}}. IEEE, \bibinfo{pages}{23--32}.
\newblock


\bibitem[\protect\citeauthoryear{Gupta, Pal, Kanade, and Shevade}{Gupta
  et~al\mbox{.}}{2017}]%
        {gupta2017deepfix}
\bibfield{author}{\bibinfo{person}{Rahul Gupta}, \bibinfo{person}{Soham Pal},
  \bibinfo{person}{Aditya Kanade}, {and} \bibinfo{person}{Shirish Shevade}.}
  \bibinfo{year}{2017}\natexlab{}.
\newblock \showarticletitle{{DeepFix}: Fixing Common C Language Errors by Deep
  Learning}. In \bibinfo{booktitle}{\emph{Proceedings of the 31st {AAAI}}}.
  \bibinfo{publisher}{{AAAI} Press}, \bibinfo{pages}{1345--1351}.
\newblock


\bibitem[\protect\citeauthoryear{Hooimeijer and Weimer}{Hooimeijer and
  Weimer}{2007}]%
        {hooimeijer2007modeling}
\bibfield{author}{\bibinfo{person}{Pieter Hooimeijer} {and}
  \bibinfo{person}{Westley Weimer}.} \bibinfo{year}{2007}\natexlab{}.
\newblock \showarticletitle{Modeling bug report quality}. In
  \bibinfo{booktitle}{\emph{Proceedings of the 22nd ASE}}. ACM,
  \bibinfo{pages}{34--43}.
\newblock


\bibitem[\protect\citeauthoryear{Hua, Zhang, Wang, and Khurshid}{Hua
  et~al\mbox{.}}{2018}]%
        {hua2018towards}
\bibfield{author}{\bibinfo{person}{Jinru Hua}, \bibinfo{person}{Mengshi Zhang},
  \bibinfo{person}{Kaiyuan Wang}, {and} \bibinfo{person}{Sarfraz Khurshid}.}
  \bibinfo{year}{2018}\natexlab{}.
\newblock \showarticletitle{Towards practical program repair with on-demand
  candidate generation}. In \bibinfo{booktitle}{\emph{Proceedings of the 40th
  ICSE}}. ACM, \bibinfo{pages}{12--23}.
\newblock


\bibitem[\protect\citeauthoryear{Jiang, Xiong, Zhang, Gao, and Chen}{Jiang
  et~al\mbox{.}}{2018}]%
        {jiang2018shaping}
\bibfield{author}{\bibinfo{person}{Jiajun Jiang}, \bibinfo{person}{Yingfei
  Xiong}, \bibinfo{person}{Hongyu Zhang}, \bibinfo{person}{Qing Gao}, {and}
  \bibinfo{person}{Xiangqun Chen}.} \bibinfo{year}{2018}\natexlab{}.
\newblock \showarticletitle{Shaping Program Repair Space with Existing Patches
  and Similar Code}. In \bibinfo{booktitle}{\emph{Proceedings of the 27th
  ISSTA}}. \bibinfo{publisher}{ACM}, \bibinfo{pages}{298--309}.
\newblock


\bibitem[\protect\citeauthoryear{Jones and Harrold}{Jones and Harrold}{2005}]%
        {jones2005empirical}
\bibfield{author}{\bibinfo{person}{James~A Jones} {and}
  \bibinfo{person}{Mary~Jean Harrold}.} \bibinfo{year}{2005}\natexlab{}.
\newblock \showarticletitle{Empirical evaluation of the tarantula automatic
  fault-localization technique}. In \bibinfo{booktitle}{\emph{Proceedings of
  the 20th ASE}}. ACM, \bibinfo{pages}{273--282}.
\newblock


\bibitem[\protect\citeauthoryear{Just, Jalali, and Ernst}{Just
  et~al\mbox{.}}{2014}]%
        {just2014defects4j}
\bibfield{author}{\bibinfo{person}{Ren{\'e} Just}, \bibinfo{person}{Darioush
  Jalali}, {and} \bibinfo{person}{Michael~D Ernst}.}
  \bibinfo{year}{2014}\natexlab{}.
\newblock \showarticletitle{{Defects4J}: {A} database of existing faults to
  enable controlled testing studies for Java programs}. In
  \bibinfo{booktitle}{\emph{Proceedings of the 23rd ISSTA}}. ACM,
  \bibinfo{pages}{437--440}.
\newblock


\bibitem[\protect\citeauthoryear{Just, Parnin, Drosos, and Ernst}{Just
  et~al\mbox{.}}{2018}]%
        {just2018comparing}
\bibfield{author}{\bibinfo{person}{Ren{\'e} Just}, \bibinfo{person}{Chris
  Parnin}, \bibinfo{person}{Ian Drosos}, {and} \bibinfo{person}{Michael~D
  Ernst}.} \bibinfo{year}{2018}\natexlab{}.
\newblock \showarticletitle{Comparing developer-provided to user-provided tests
  for fault localization and automated program repair}. In
  \bibinfo{booktitle}{\emph{Proceedings of the 27th ACM SIGSOFT International
  Symposium on Software Testing and Analysis}}. ACM, \bibinfo{pages}{287--297}.
\newblock


\bibitem[\protect\citeauthoryear{Juzgado, Moreno, and Strigel}{Juzgado
  et~al\mbox{.}}{2006}]%
        {juristo2006guest}
\bibfield{author}{\bibinfo{person}{Natalia~Juristo Juzgado},
  \bibinfo{person}{Ana~Mar{\'{\i}}a Moreno}, {and} \bibinfo{person}{Wolfgang
  Strigel}.} \bibinfo{year}{2006}\natexlab{}.
\newblock \showarticletitle{Guest editors' introduction: Software testing
  practices in industry}.
\newblock \bibinfo{journal}{\emph{IEEE Software}} \bibinfo{volume}{23},
  \bibinfo{number}{4} (\bibinfo{year}{2006}), \bibinfo{pages}{19--21}.
\newblock


\bibitem[\protect\citeauthoryear{Karaa and Grib{\^a}a}{Karaa and
  Grib{\^a}a}{2013}]%
        {karaa2013information}
\bibfield{author}{\bibinfo{person}{Wahiba Ben~Abdessalem Karaa} {and}
  \bibinfo{person}{Nidhal Grib{\^a}a}.} \bibinfo{year}{2013}\natexlab{}.
\newblock \showarticletitle{Information retrieval with porter stemmer: a new
  version for English}.
\newblock In \bibinfo{booktitle}{\emph{Advances in computational science,
  engineering and information technology}}. \bibinfo{publisher}{Springer},
  \bibinfo{pages}{243--254}.
\newblock


\bibitem[\protect\citeauthoryear{Ke, Stolee, Le~Goues, and Brun}{Ke
  et~al\mbox{.}}{2015}]%
        {ke2015repairing}
\bibfield{author}{\bibinfo{person}{Yalin Ke}, \bibinfo{person}{Kathryn~T
  Stolee}, \bibinfo{person}{Claire Le~Goues}, {and} \bibinfo{person}{Yuriy
  Brun}.} \bibinfo{year}{2015}\natexlab{}.
\newblock \showarticletitle{Repairing programs with semantic code search (t)}.
  In \bibinfo{booktitle}{\emph{Proceedings of the 30th ASE}}.
  \bibinfo{publisher}{{IEEE}}, \bibinfo{pages}{295--306}.
\newblock


\bibitem[\protect\citeauthoryear{Kim, Nam, Song, and Kim}{Kim
  et~al\mbox{.}}{2013}]%
        {kim2013automatic}
\bibfield{author}{\bibinfo{person}{Dongsun Kim}, \bibinfo{person}{Jaechang
  Nam}, \bibinfo{person}{Jaewoo Song}, {and} \bibinfo{person}{Sunghun Kim}.}
  \bibinfo{year}{2013}\natexlab{}.
\newblock \showarticletitle{Automatic patch generation learned from
  human-written patches}. In \bibinfo{booktitle}{\emph{Proceedings of the 35th
  ICSE}}. \bibinfo{publisher}{{IEEE}}, \bibinfo{pages}{802--811}.
\newblock


\bibitem[\protect\citeauthoryear{Kochhar, Bissyand{\'e}, Lo, and Jiang}{Kochhar
  et~al\mbox{.}}{2013}]%
        {kochhar2013empirical}
\bibfield{author}{\bibinfo{person}{Pavneet~Singh Kochhar},
  \bibinfo{person}{Tegawend{\'e}~F Bissyand{\'e}}, \bibinfo{person}{David Lo},
  {and} \bibinfo{person}{Lingxiao Jiang}.} \bibinfo{year}{2013}\natexlab{}.
\newblock \showarticletitle{An empirical study of adoption of software testing
  in open source projects}. In \bibinfo{booktitle}{\emph{Proceedings of the
  13th QRS}}. IEEE, \bibinfo{pages}{103--112}.
\newblock


\bibitem[\protect\citeauthoryear{Kolmogorov and Fomin}{Kolmogorov and
  Fomin}{1999}]%
        {kolmogorov1999elements}
\bibfield{author}{\bibinfo{person}{A.~N. Kolmogorov} {and}
  \bibinfo{person}{S.~V. Fomin}.} \bibinfo{year}{1999}\natexlab{}.
\newblock \bibinfo{booktitle}{\emph{Elements of the {{Theory}} of {{Functions}}
  and {{Functional Analysis}}} (\bibinfo{edition}{dover books on mathematics
  edition} ed.)}.
\newblock \bibinfo{publisher}{{Dover Publications}}, \bibinfo{address}{Mineola,
  NY}.
\newblock


\bibitem[\protect\citeauthoryear{Koyuncu, Bissyand{\'e}, Kim, Klein, Monperrus,
  and Le~Traon}{Koyuncu et~al\mbox{.}}{2017}]%
        {koyuncu2017impact}
\bibfield{author}{\bibinfo{person}{Anil Koyuncu},
  \bibinfo{person}{Tegawend{\'e}~F Bissyand{\'e}}, \bibinfo{person}{Dongsun
  Kim}, \bibinfo{person}{Jacques Klein}, \bibinfo{person}{Martin Monperrus},
  {and} \bibinfo{person}{Yves Le~Traon}.} \bibinfo{year}{2017}\natexlab{}.
\newblock \showarticletitle{Impact of tool support in patch construction}. In
  \bibinfo{booktitle}{\emph{Proceedings of the 26th ISSTA}}. ACM,
  \bibinfo{pages}{237--248}.
\newblock


\bibitem[\protect\citeauthoryear{Koyuncu, Bissyand{\'e}, Kim, Liu, Klein,
  Monperrus, and Traon}{Koyuncu et~al\mbox{.}}{2019}]%
        {koyuncu2019d}
\bibfield{author}{\bibinfo{person}{Anil Koyuncu},
  \bibinfo{person}{Tegawend{\'e}~F Bissyand{\'e}}, \bibinfo{person}{Dongsun
  Kim}, \bibinfo{person}{Kui Liu}, \bibinfo{person}{Jacques Klein},
  \bibinfo{person}{Martin Monperrus}, {and} \bibinfo{person}{Yves~Le Traon}.}
  \bibinfo{year}{2019}\natexlab{}.
\newblock \showarticletitle{D\&C: A Divide-and-Conquer Approach to IR-based Bug
  Localization}.
\newblock \bibinfo{journal}{\emph{arXiv preprint arXiv:1902.02703}}
  (\bibinfo{year}{2019}).
\newblock


\bibitem[\protect\citeauthoryear{Koyuncu, Liu, F.~Bissyand\'e, Kim, Klein,
  Monperrus, and Le~Traon}{Koyuncu et~al\mbox{.}}{2018}]%
        {koyuncu2018fixminer}
\bibfield{author}{\bibinfo{person}{Anil Koyuncu}, \bibinfo{person}{Kui Liu},
  \bibinfo{person}{Tegawend\'e F.~Bissyand\'e}, \bibinfo{person}{Dongsun Kim},
  \bibinfo{person}{Jacques Klein}, \bibinfo{person}{Martin Monperrus}, {and}
  \bibinfo{person}{Yves Le~Traon}.} \bibinfo{year}{2018}\natexlab{}.
\newblock \showarticletitle{FixMiner: Mining Relevant Fix Patterns for
  Automated Program Repair}.
\newblock \bibinfo{journal}{\emph{arXiv preprint arXiv:1810.01791}}
  (\bibinfo{year}{2018}).
\newblock


\bibitem[\protect\citeauthoryear{Le, Chu, Lo, Le~Goues, and Visser}{Le
  et~al\mbox{.}}{2017}]%
        {le2017s3}
\bibfield{author}{\bibinfo{person}{Xuan-Bach~D Le}, \bibinfo{person}{Duc-Hiep
  Chu}, \bibinfo{person}{David Lo}, \bibinfo{person}{Claire Le~Goues}, {and}
  \bibinfo{person}{Willem Visser}.} \bibinfo{year}{2017}\natexlab{}.
\newblock \showarticletitle{S3: syntax-and semantic-guided repair synthesis via
  programming by examples}. In \bibinfo{booktitle}{\emph{Proceedings of the
  11th FSE}}. \bibinfo{publisher}{ACM}, \bibinfo{pages}{593--604}.
\newblock


\bibitem[\protect\citeauthoryear{Le, Le, Lo, and Le~Goues}{Le
  et~al\mbox{.}}{2016a}]%
        {le2016enhancing}
\bibfield{author}{\bibinfo{person}{Xuan-Bach~D Le}, \bibinfo{person}{Quang~Loc
  Le}, \bibinfo{person}{David Lo}, {and} \bibinfo{person}{Claire Le~Goues}.}
  \bibinfo{year}{2016}\natexlab{a}.
\newblock \showarticletitle{Enhancing automated program repair with deductive
  verification}. In \bibinfo{booktitle}{\emph{Proceedings of the 32nd ICSME}}.
  \bibinfo{publisher}{{IEEE}}, \bibinfo{pages}{428--432}.
\newblock


\bibitem[\protect\citeauthoryear{Le, Thung, Lo, and Le~Goues}{Le
  et~al\mbox{.}}{2018}]%
        {le2018overfitting}
\bibfield{author}{\bibinfo{person}{Xuan Bach~D Le}, \bibinfo{person}{Ferdian
  Thung}, \bibinfo{person}{David Lo}, {and} \bibinfo{person}{Claire Le~Goues}.}
  \bibinfo{year}{2018}\natexlab{}.
\newblock \showarticletitle{Overfitting in semantics-based automated program
  repair}.
\newblock \bibinfo{journal}{\emph{EMSE Journal}} (\bibinfo{year}{2018}),
  \bibinfo{pages}{1--27}.
\newblock


\bibitem[\protect\citeauthoryear{Le, Bao, Lo, Xia, and Li}{Le
  et~al\mbox{.}}{2019}]%
        {le2019reliability}
\bibfield{author}{\bibinfo{person}{Xuan{-}Bach~D. Le},
  \bibinfo{person}{Lingfeng Bao}, \bibinfo{person}{David Lo},
  \bibinfo{person}{Xin Xia}, {and} \bibinfo{person}{Shanping Li}.}
  \bibinfo{year}{2019}\natexlab{}.
\newblock \showarticletitle{On Reliability of Patch Correctness Assessment}. In
  \bibinfo{booktitle}{\emph{Proceedings of the 41st ICSE}}.
\newblock


\bibitem[\protect\citeauthoryear{Le, Lo, and {Le Goues}}{Le
  et~al\mbox{.}}{2016b}]%
        {le2016history}
\bibfield{author}{\bibinfo{person}{Xuan{-}Bach~D. Le}, \bibinfo{person}{David
  Lo}, {and} \bibinfo{person}{Claire {Le Goues}}.}
  \bibinfo{year}{2016}\natexlab{b}.
\newblock \showarticletitle{History Driven Program Repair}. In
  \bibinfo{booktitle}{\emph{Proceedings of the 23rd SANER}},
  Vol.~\bibinfo{volume}{1}. \bibinfo{publisher}{IEEE},
  \bibinfo{pages}{213--224}.
\newblock


\bibitem[\protect\citeauthoryear{Le~Goues, Dewey-Vogt, Forrest, and
  Weimer}{Le~Goues et~al\mbox{.}}{2012a}]%
        {le2012systematic}
\bibfield{author}{\bibinfo{person}{Claire Le~Goues}, \bibinfo{person}{Michael
  Dewey-Vogt}, \bibinfo{person}{Stephanie Forrest}, {and}
  \bibinfo{person}{Westley Weimer}.} \bibinfo{year}{2012}\natexlab{a}.
\newblock \showarticletitle{A systematic study of automated program repair:
  Fixing 55 out of 105 bugs for \$8 each}. In
  \bibinfo{booktitle}{\emph{Proceedings of the 34th ICSE}}. IEEE,
  \bibinfo{pages}{3--13}.
\newblock


\bibitem[\protect\citeauthoryear{Le~Goues, Nguyen, Forrest, and
  Weimer}{Le~Goues et~al\mbox{.}}{2012b}]%
        {le2012genprog}
\bibfield{author}{\bibinfo{person}{Claire Le~Goues}, \bibinfo{person}{ThanhVu
  Nguyen}, \bibinfo{person}{Stephanie Forrest}, {and} \bibinfo{person}{Westley
  Weimer}.} \bibinfo{year}{2012}\natexlab{b}.
\newblock \showarticletitle{{GenProg}: A generic method for automatic software
  repair}.
\newblock \bibinfo{journal}{\emph{TSE}} \bibinfo{volume}{38},
  \bibinfo{number}{1} (\bibinfo{year}{2012}), \bibinfo{pages}{54--72}.
\newblock


\bibitem[\protect\citeauthoryear{{Le Goues}, Nguyen, Forrest, and Weimer}{{Le
  Goues} et~al\mbox{.}}{2012}]%
        {claire2012genprog}
\bibfield{author}{\bibinfo{person}{Claire {Le Goues}}, \bibinfo{person}{ThanhVu
  Nguyen}, \bibinfo{person}{Stephanie Forrest}, {and} \bibinfo{person}{Westley
  Weimer}.} \bibinfo{year}{2012}\natexlab{}.
\newblock \showarticletitle{{GenProg}: {A} Generic Method for Automatic
  Software Repair}.
\newblock \bibinfo{journal}{\emph{TSE}} \bibinfo{volume}{38},
  \bibinfo{number}{1} (\bibinfo{year}{2012}), \bibinfo{pages}{54--72}.
\newblock


\bibitem[\protect\citeauthoryear{Le~Goues and Weimer}{Le~Goues and
  Weimer}{2009}]%
        {le2009specification}
\bibfield{author}{\bibinfo{person}{Claire Le~Goues} {and}
  \bibinfo{person}{Westley Weimer}.} \bibinfo{year}{2009}\natexlab{}.
\newblock \showarticletitle{Specification mining with few false positives}. In
  \bibinfo{booktitle}{\emph{Proceedings of the 15th TACAS}}. Springer,
  \bibinfo{pages}{292--306}.
\newblock


\bibitem[\protect\citeauthoryear{Lee, Kim, Bissyand{\'e}, Jung, and
  Le~Traon}{Lee et~al\mbox{.}}{2018}]%
        {lee2018bench4bl}
\bibfield{author}{\bibinfo{person}{Jaekwon Lee}, \bibinfo{person}{Dongsun Kim},
  \bibinfo{person}{Tegawend{\'e}~F Bissyand{\'e}}, \bibinfo{person}{Woosung
  Jung}, {and} \bibinfo{person}{Yves Le~Traon}.}
  \bibinfo{year}{2018}\natexlab{}.
\newblock \showarticletitle{Bench4bl: reproducibility study on the performance
  of ir-based bug localization}. In \bibinfo{booktitle}{\emph{Proceedings of
  the 27th ISSTA}}. ACM, \bibinfo{pages}{61--72}.
\newblock


\bibitem[\protect\citeauthoryear{Liblit, Naik, Zheng, Aiken, and Jordan}{Liblit
  et~al\mbox{.}}{2005}]%
        {liblit2005scalable}
\bibfield{author}{\bibinfo{person}{Ben Liblit}, \bibinfo{person}{Mayur Naik},
  \bibinfo{person}{Alice~X Zheng}, \bibinfo{person}{Alex Aiken}, {and}
  \bibinfo{person}{Michael~I Jordan}.} \bibinfo{year}{2005}\natexlab{}.
\newblock \showarticletitle{Scalable statistical bug isolation}. In
  \bibinfo{booktitle}{\emph{Proceedings of the 26th PLDI}}.
  \bibinfo{publisher}{ACM}, \bibinfo{pages}{15--26}.
\newblock


\bibitem[\protect\citeauthoryear{Liu, Yang, Tan, and Hafiz}{Liu
  et~al\mbox{.}}{2013}]%
        {liu2013r2fix}
\bibfield{author}{\bibinfo{person}{Chen Liu}, \bibinfo{person}{Jinqiu Yang},
  \bibinfo{person}{Lin Tan}, {and} \bibinfo{person}{Munawar Hafiz}.}
  \bibinfo{year}{2013}\natexlab{}.
\newblock \showarticletitle{{R2Fix}: Automatically generating bug fixes from
  bug reports}. In \bibinfo{booktitle}{\emph{Proceedings of the 6th ICST}}.
  IEEE, \bibinfo{pages}{282--291}.
\newblock


\bibitem[\protect\citeauthoryear{Liu, Anil, Kim, Kim, and Bissyand\'e}{Liu
  et~al\mbox{.}}{2018a}]%
        {liu2018lsrepair}
\bibfield{author}{\bibinfo{person}{Kui Liu}, \bibinfo{person}{Koyuncu Anil},
  \bibinfo{person}{Kisub Kim}, \bibinfo{person}{Dongsun Kim}, {and}
  \bibinfo{person}{Tegawend\'e~F. Bissyand\'e}.}
  \bibinfo{year}{2018}\natexlab{a}.
\newblock \showarticletitle{{LSRepair}: Live Search of Fix Ingredients for
  Automated Program Repair}. In \bibinfo{booktitle}{\emph{Proceedings of the
  25th APSEC}}. \bibinfo{publisher}{IEEE}, \bibinfo{pages}{658--662}.
\newblock


\bibitem[\protect\citeauthoryear{Liu, Kim, Bissyand{\'e}, Kim, Kim, Koyuncu,
  Kim, and Le~Traon}{Liu et~al\mbox{.}}{2019a}]%
        {liu2019learning}
\bibfield{author}{\bibinfo{person}{Kui Liu}, \bibinfo{person}{Dongsun Kim},
  \bibinfo{person}{Tegawend{\'e}~F. Bissyand{\'e}}, \bibinfo{person}{Taeyoung
  Kim}, \bibinfo{person}{Kisub Kim}, \bibinfo{person}{Anil Koyuncu},
  \bibinfo{person}{Suntae Kim}, {and} \bibinfo{person}{Yves Le~Traon}.}
  \bibinfo{year}{2019}\natexlab{a}.
\newblock \showarticletitle{Learning to Spot and Refactor Inconsistent Method
  Names}. In \bibinfo{booktitle}{\emph{Proceedings of the 41st ICSE}}. IEEE,
  \bibinfo{pages}{1--12}.
\newblock


\bibitem[\protect\citeauthoryear{Liu, Kim, Bissyand{\'e}, Yoo, and
  Le~Traon}{Liu et~al\mbox{.}}{2018b}]%
        {liu2018mining2}
\bibfield{author}{\bibinfo{person}{Kui Liu}, \bibinfo{person}{Dongsun Kim},
  \bibinfo{person}{Tegawend{\'e}~F Bissyand{\'e}}, \bibinfo{person}{Shin Yoo},
  {and} \bibinfo{person}{Yves Le~Traon}.} \bibinfo{year}{2018}\natexlab{b}.
\newblock \showarticletitle{Mining fix patterns for findbugs violations}.
\newblock \bibinfo{journal}{\emph{TSE}} (\bibinfo{year}{2018}).
\newblock


\bibitem[\protect\citeauthoryear{Liu, Kim, Koyuncu, Li, Bissyand{\'e}, and
  Le~Traon}{Liu et~al\mbox{.}}{2018c}]%
        {liu2018closer}
\bibfield{author}{\bibinfo{person}{Kui Liu}, \bibinfo{person}{Dongsun Kim},
  \bibinfo{person}{Anil Koyuncu}, \bibinfo{person}{Li Li},
  \bibinfo{person}{Tegawend{\'e}~F Bissyand{\'e}}, {and} \bibinfo{person}{Yves
  Le~Traon}.} \bibinfo{year}{2018}\natexlab{c}.
\newblock \showarticletitle{A closer look at real-world patches}. In
  \bibinfo{booktitle}{\emph{Proceedings of the 34th ICSME}}. IEEE,
  \bibinfo{pages}{275--286}.
\newblock


\bibitem[\protect\citeauthoryear{Liu, Koyuncu, Bissyand{\'e}, Kim, Klein, and
  Le~Traon}{Liu et~al\mbox{.}}{2019b}]%
        {liu2019you}
\bibfield{author}{\bibinfo{person}{Kui Liu}, \bibinfo{person}{Anil Koyuncu},
  \bibinfo{person}{Tegawend{\'e}~F. Bissyand{\'e}}, \bibinfo{person}{Dongsun
  Kim}, \bibinfo{person}{Jacques Klein}, {and} \bibinfo{person}{Yves
  Le~Traon}.} \bibinfo{year}{2019}\natexlab{b}.
\newblock \showarticletitle{You Cannot Fix What You Cannot Find! An
  Investigation of Fault Localization Bias in Benchmarking Automated Program
  Repair Systems}. In \bibinfo{booktitle}{\emph{Proceedings of the 12th ICST}}.
  IEEE, \bibinfo{pages}{102--113}.
\newblock


\bibitem[\protect\citeauthoryear{Liu, Koyuncu, Kim, and Bissyand{\'e}}{Liu
  et~al\mbox{.}}{2019c}]%
        {liu2019avatar}
\bibfield{author}{\bibinfo{person}{Kui Liu}, \bibinfo{person}{Anil Koyuncu},
  \bibinfo{person}{Dongsun Kim}, {and} \bibinfo{person}{Tegawend{\'e}~F.
  Bissyand{\'e}}.} \bibinfo{year}{2019}\natexlab{c}.
\newblock \showarticletitle{{AVATAR}: Fixing Semantic Bugs with Fix Patterns of
  Static Analysis Violations}. In \bibinfo{booktitle}{\emph{Proceedings of the
  26th SANER}}. IEEE, \bibinfo{pages}{1--12}.
\newblock


\bibitem[\protect\citeauthoryear{Liu, Koyuncu, Kim, and Bissyand{\'e}}{Liu
  et~al\mbox{.}}{2019d}]%
        {liu2019tbar}
\bibfield{author}{\bibinfo{person}{Kui Liu}, \bibinfo{person}{Anil Koyuncu},
  \bibinfo{person}{Dongsun Kim}, {and} \bibinfo{person}{Tegawend{\'e}~F.
  Bissyand{\'e}}.} \bibinfo{year}{2019}\natexlab{d}.
\newblock \showarticletitle{{TBar} : Revisiting Template-based Automated
  Program Repair}. In \bibinfo{booktitle}{\emph{Proceedings of the 28th
  ISSTA}}. ACM.
\newblock


\bibitem[\protect\citeauthoryear{Liu and Zhong}{Liu and Zhong}{2018}]%
        {liu2018mining}
\bibfield{author}{\bibinfo{person}{Xuliang Liu} {and} \bibinfo{person}{Hao
  Zhong}.} \bibinfo{year}{2018}\natexlab{}.
\newblock \showarticletitle{Mining stackoverflow for program repair}. In
  \bibinfo{booktitle}{\emph{Proceedings of the 25th SANER}}. IEEE,
  \bibinfo{pages}{118--129}.
\newblock


\bibitem[\protect\citeauthoryear{Long, Amidon, and Rinard}{Long
  et~al\mbox{.}}{2017}]%
        {long2017automatic}
\bibfield{author}{\bibinfo{person}{Fan Long}, \bibinfo{person}{Peter Amidon},
  {and} \bibinfo{person}{Martin Rinard}.} \bibinfo{year}{2017}\natexlab{}.
\newblock \showarticletitle{Automatic inference of code transforms for patch
  generation}. In \bibinfo{booktitle}{\emph{Proceedings of the 11th FSE}}.
  \bibinfo{publisher}{ACM}, \bibinfo{pages}{727--739}.
\newblock


\bibitem[\protect\citeauthoryear{Long and Rinard}{Long and Rinard}{2015}]%
        {long2015staged}
\bibfield{author}{\bibinfo{person}{Fan Long} {and} \bibinfo{person}{Martin
  Rinard}.} \bibinfo{year}{2015}\natexlab{}.
\newblock \showarticletitle{Staged program repair with condition synthesis}. In
  \bibinfo{booktitle}{\emph{Proceedings of the 10th FSE}}.
  \bibinfo{publisher}{ACM}, \bibinfo{pages}{166--178}.
\newblock


\bibitem[\protect\citeauthoryear{Long and Rinard}{Long and Rinard}{2016a}]%
        {long2016analysis}
\bibfield{author}{\bibinfo{person}{Fan Long} {and} \bibinfo{person}{Martin
  Rinard}.} \bibinfo{year}{2016}\natexlab{a}.
\newblock \showarticletitle{An analysis of the search spaces for generate and
  validate patch generation systems}. In \bibinfo{booktitle}{\emph{Proceedings
  of the 38th ICSE}}. ACM, \bibinfo{pages}{702--713}.
\newblock


\bibitem[\protect\citeauthoryear{Long and Rinard}{Long and Rinard}{2016b}]%
        {long2016automatic}
\bibfield{author}{\bibinfo{person}{Fan Long} {and} \bibinfo{person}{Martin
  Rinard}.} \bibinfo{year}{2016}\natexlab{b}.
\newblock \showarticletitle{Automatic patch generation by learning correct
  code}. In \bibinfo{booktitle}{\emph{Proceedings of the 43rd POPL}}.
  \bibinfo{publisher}{ACM}, \bibinfo{pages}{298--312}.
\newblock


\bibitem[\protect\citeauthoryear{{LUCIA}, Thung, Lo, and Jiang}{{LUCIA}
  et~al\mbox{.}}{2012}]%
        {thung2012faults}
\bibfield{author}{\bibinfo{person}{Lucia {LUCIA}}, \bibinfo{person}{Ferdian
  Thung}, \bibinfo{person}{David Lo}, {and} \bibinfo{person}{Lingxiao Jiang}.}
  \bibinfo{year}{2012}\natexlab{}.
\newblock \showarticletitle{Are Faults Localizable?}. In
  \bibinfo{booktitle}{\emph{Proceedings of the 9th MSR}}.
  \bibinfo{pages}{74--77}.
\newblock


\bibitem[\protect\citeauthoryear{Lukins, Kraft, and Etzkorn}{Lukins
  et~al\mbox{.}}{2010}]%
        {lukins2010bug}
\bibfield{author}{\bibinfo{person}{Stacy~K Lukins}, \bibinfo{person}{Nicholas~A
  Kraft}, {and} \bibinfo{person}{Letha~H Etzkorn}.}
  \bibinfo{year}{2010}\natexlab{}.
\newblock \showarticletitle{Bug localization using latent {Dirichlet}
  allocation}.
\newblock \bibinfo{journal}{\emph{IST}} \bibinfo{volume}{52},
  \bibinfo{number}{9} (\bibinfo{year}{2010}), \bibinfo{pages}{972--990}.
\newblock


\bibitem[\protect\citeauthoryear{Mao, Lei, Dai, Qi, and Wang}{Mao
  et~al\mbox{.}}{2014}]%
        {mao2014slice}
\bibfield{author}{\bibinfo{person}{Xiaoguang Mao}, \bibinfo{person}{Yan Lei},
  \bibinfo{person}{Ziying Dai}, \bibinfo{person}{Yuhua Qi}, {and}
  \bibinfo{person}{Chengsong Wang}.} \bibinfo{year}{2014}\natexlab{}.
\newblock \showarticletitle{Slice-based statistical fault localization}.
\newblock \bibinfo{journal}{\emph{JSS}}  \bibinfo{volume}{89}
  (\bibinfo{year}{2014}), \bibinfo{pages}{51--62}.
\newblock


\bibitem[\protect\citeauthoryear{Martinez and Monperrus}{Martinez and
  Monperrus}{2016}]%
        {martinez2016astor}
\bibfield{author}{\bibinfo{person}{Matias Martinez} {and}
  \bibinfo{person}{Martin Monperrus}.} \bibinfo{year}{2016}\natexlab{}.
\newblock \showarticletitle{Astor: A program repair library for java}. In
  \bibinfo{booktitle}{\emph{Proceedings of the 25th ISSTA}}. ACM,
  \bibinfo{pages}{441--444}.
\newblock


\bibitem[\protect\citeauthoryear{Martinez and Monperrus}{Martinez and
  Monperrus}{2018}]%
        {martinez2018ultra}
\bibfield{author}{\bibinfo{person}{Matias Martinez} {and}
  \bibinfo{person}{Martin Monperrus}.} \bibinfo{year}{2018}\natexlab{}.
\newblock \showarticletitle{Ultra-Large Repair Search Space with Automatically
  Mined Templates: The Cardumen Mode of Astor}. In
  \bibinfo{booktitle}{\emph{Proceedings of the 10th SSBSE}}. Springer,
  \bibinfo{pages}{65--86}.
\newblock


\bibitem[\protect\citeauthoryear{Mechtaev, Nguyen, Noller, Grunske, and
  Roychoudhury}{Mechtaev et~al\mbox{.}}{2018}]%
        {mechtaev2018semantic}
\bibfield{author}{\bibinfo{person}{Sergey Mechtaev}, \bibinfo{person}{Manh-Dung
  Nguyen}, \bibinfo{person}{Yannic Noller}, \bibinfo{person}{Lars Grunske},
  {and} \bibinfo{person}{Abhik Roychoudhury}.} \bibinfo{year}{2018}\natexlab{}.
\newblock \showarticletitle{Semantic Program Repair Using a Reference
  Implementation}. In \bibinfo{booktitle}{\emph{Proceedings of the 40th ICSE}}.
  \bibinfo{publisher}{{ACM}}, \bibinfo{pages}{298--309}.
\newblock


\bibitem[\protect\citeauthoryear{Mechtaev, Yi, and Roychoudhury}{Mechtaev
  et~al\mbox{.}}{2015}]%
        {mechtaev2015directfix}
\bibfield{author}{\bibinfo{person}{Sergey Mechtaev}, \bibinfo{person}{Jooyong
  Yi}, {and} \bibinfo{person}{Abhik Roychoudhury}.}
  \bibinfo{year}{2015}\natexlab{}.
\newblock \showarticletitle{Directfix: Looking for simple program repairs}. In
  \bibinfo{booktitle}{\emph{Proceedings of the 37th ICSE}}.
  \bibinfo{publisher}{IEEE Press}, \bibinfo{pages}{448--458}.
\newblock


\bibitem[\protect\citeauthoryear{Monperrus}{Monperrus}{2014}]%
        {monperrus2014critical}
\bibfield{author}{\bibinfo{person}{Martin Monperrus}.}
  \bibinfo{year}{2014}\natexlab{}.
\newblock \showarticletitle{A critical review of automatic patch generation
  learned from human-written patches: essay on the problem statement and the
  evaluation of automatic software repair}. In
  \bibinfo{booktitle}{\emph{Proceedings of the 36th ICSE}}. ACM,
  \bibinfo{pages}{234--242}.
\newblock


\bibitem[\protect\citeauthoryear{Nayrolles and Hamou-Lhadj}{Nayrolles and
  Hamou-Lhadj}{2018}]%
        {nayrolles2018clever}
\bibfield{author}{\bibinfo{person}{Mathieu Nayrolles} {and}
  \bibinfo{person}{Abdelwahab Hamou-Lhadj}.} \bibinfo{year}{2018}\natexlab{}.
\newblock \showarticletitle{CLEVER: combining code metrics with clone detection
  for just-in-time fault prevention and resolution in large industrial
  projects}. In \bibinfo{booktitle}{\emph{Proceedings of the 15th MSR}}. ACM,
  \bibinfo{pages}{153--164}.
\newblock


\bibitem[\protect\citeauthoryear{Nguyen, Qi, Roychoudhury, and Chandra}{Nguyen
  et~al\mbox{.}}{2013}]%
        {nguyen2013semfix}
\bibfield{author}{\bibinfo{person}{Hoang Duong~Thien Nguyen},
  \bibinfo{person}{Dawei Qi}, \bibinfo{person}{Abhik Roychoudhury}, {and}
  \bibinfo{person}{Satish Chandra}.} \bibinfo{year}{2013}\natexlab{}.
\newblock \showarticletitle{{SemFix}: program repair via semantic analysis}. In
  \bibinfo{booktitle}{\emph{Proceedings of the 35th ICSE}}.
  \bibinfo{publisher}{IEEE}, \bibinfo{pages}{772--781}.
\newblock


\bibitem[\protect\citeauthoryear{Padioleau, Lawall, Hansen, and
  Muller}{Padioleau et~al\mbox{.}}{2008}]%
        {padioleau2008documenting}
\bibfield{author}{\bibinfo{person}{Yoann Padioleau}, \bibinfo{person}{Julia
  Lawall}, \bibinfo{person}{Ren{\'e}~Rydhof Hansen}, {and}
  \bibinfo{person}{Gilles Muller}.} \bibinfo{year}{2008}\natexlab{}.
\newblock \showarticletitle{Documenting and automating collateral evolutions in
  Linux device drivers}. In \bibinfo{booktitle}{\emph{Proceedings of 3rd
  EuroSys}}, Vol.~\bibinfo{volume}{42}. ACM, \bibinfo{pages}{247--260}.
\newblock


\bibitem[\protect\citeauthoryear{Parnin and Orso}{Parnin and Orso}{2011}]%
        {parnin2011automated}
\bibfield{author}{\bibinfo{person}{Chris Parnin} {and}
  \bibinfo{person}{Alessandro Orso}.} \bibinfo{year}{2011}\natexlab{}.
\newblock \showarticletitle{Are automated debugging techniques actually helping
  programmers?}. In \bibinfo{booktitle}{\emph{Proceedings of the 20th ISSTA}}.
  ACM, \bibinfo{pages}{199--209}.
\newblock


\bibitem[\protect\citeauthoryear{Petri{\'c}, Hall, and Bowes}{Petri{\'c}
  et~al\mbox{.}}{2018}]%
        {petric2018effectively}
\bibfield{author}{\bibinfo{person}{Jean Petri{\'c}}, \bibinfo{person}{Tracy
  Hall}, {and} \bibinfo{person}{David Bowes}.} \bibinfo{year}{2018}\natexlab{}.
\newblock \showarticletitle{How Effectively Is Defective Code Actually Tested?:
  An Analysis of JUnit Tests in Seven Open Source Systems}. In
  \bibinfo{booktitle}{\emph{Proceedings of the 14th PROMISE}}. ACM,
  \bibinfo{pages}{42--51}.
\newblock


\bibitem[\protect\citeauthoryear{Qi, Long, Achour, and Rinard}{Qi
  et~al\mbox{.}}{2015}]%
        {qi2015analysis}
\bibfield{author}{\bibinfo{person}{Zichao Qi}, \bibinfo{person}{Fan Long},
  \bibinfo{person}{Sara Achour}, {and} \bibinfo{person}{Martin Rinard}.}
  \bibinfo{year}{2015}\natexlab{}.
\newblock \showarticletitle{An analysis of patch plausibility and correctness
  for generate-and-validate patch generation systems}. In
  \bibinfo{booktitle}{\emph{Proceedings of the 24th ISSTA}}. ACM,
  \bibinfo{pages}{24--36}.
\newblock


\bibitem[\protect\citeauthoryear{Rolim, Soares, D'Antoni, Polozov, Gulwani,
  Gheyi, Suzuki, and Hartmann}{Rolim et~al\mbox{.}}{2017}]%
        {rolim2017learning}
\bibfield{author}{\bibinfo{person}{Reudismam Rolim}, \bibinfo{person}{Gustavo
  Soares}, \bibinfo{person}{Loris D'Antoni}, \bibinfo{person}{Oleksandr
  Polozov}, \bibinfo{person}{Sumit Gulwani}, \bibinfo{person}{Rohit Gheyi},
  \bibinfo{person}{Ryo Suzuki}, {and} \bibinfo{person}{Bj{\"o}rn Hartmann}.}
  \bibinfo{year}{2017}\natexlab{}.
\newblock \showarticletitle{Learning syntactic program transformations from
  examples}. In \bibinfo{booktitle}{\emph{Proceedings of the 39th ICSE}}.
  IEEE/ACM, \bibinfo{pages}{404--415}.
\newblock


\bibitem[\protect\citeauthoryear{Saha, Lyu, Lam, Yoshida, and Prasad}{Saha
  et~al\mbox{.}}{2018}]%
        {saha2018bugs}
\bibfield{author}{\bibinfo{person}{Ripon Saha}, \bibinfo{person}{Yingjun Lyu},
  \bibinfo{person}{Wing Lam}, \bibinfo{person}{Hiroaki Yoshida}, {and}
  \bibinfo{person}{Mukul Prasad}.} \bibinfo{year}{2018}\natexlab{}.
\newblock \showarticletitle{Bugs.jar: a large-scale, diverse dataset of
  real-world java bugs}. In \bibinfo{booktitle}{\emph{Proceedings of the 15th
  MSR}}. IEEE, \bibinfo{pages}{10--13}.
\newblock


\bibitem[\protect\citeauthoryear{Saha, Lease, Khurshid, and Perry}{Saha
  et~al\mbox{.}}{2013}]%
        {saha2013improving}
\bibfield{author}{\bibinfo{person}{Ripon~K Saha}, \bibinfo{person}{Matthew
  Lease}, \bibinfo{person}{Sarfraz Khurshid}, {and} \bibinfo{person}{Dewayne~E
  Perry}.} \bibinfo{year}{2013}\natexlab{}.
\newblock \showarticletitle{Improving bug localization using structured
  information retrieval}. In \bibinfo{booktitle}{\emph{Proceedings of the 28th
  ASE}}. IEEE, \bibinfo{pages}{345--355}.
\newblock


\bibitem[\protect\citeauthoryear{Saha, Lyu, Yoshida, and Prasad}{Saha
  et~al\mbox{.}}{2017}]%
        {saha2017elixir}
\bibfield{author}{\bibinfo{person}{Ripon~K Saha}, \bibinfo{person}{Yingjun
  Lyu}, \bibinfo{person}{Hiroaki Yoshida}, {and} \bibinfo{person}{Mukul~R
  Prasad}.} \bibinfo{year}{2017}\natexlab{}.
\newblock \showarticletitle{{ELIXIR}: Effective object-oriented program
  repair}. In \bibinfo{booktitle}{\emph{Proceedings of the 32nd ASE}}. IEEE,
  \bibinfo{pages}{648--659}.
\newblock


\bibitem[\protect\citeauthoryear{Schulte, Fry, Fast, Weimer, and
  Forrest}{Schulte et~al\mbox{.}}{2014}]%
        {schulte2014software}
\bibfield{author}{\bibinfo{person}{Eric Schulte}, \bibinfo{person}{Zachary~P
  Fry}, \bibinfo{person}{Ethan Fast}, \bibinfo{person}{Westley Weimer}, {and}
  \bibinfo{person}{Stephanie Forrest}.} \bibinfo{year}{2014}\natexlab{}.
\newblock \showarticletitle{Software mutational robustness}.
\newblock \bibinfo{journal}{\emph{Genetic Programming and Evolvable Machines}}
  \bibinfo{volume}{15}, \bibinfo{number}{3} (\bibinfo{year}{2014}),
  \bibinfo{pages}{281--312}.
\newblock


\bibitem[\protect\citeauthoryear{Scott, Bader, and Chandra}{Scott
  et~al\mbox{.}}{2019}]%
        {getafix2019}
\bibfield{author}{\bibinfo{person}{Andrew Scott}, \bibinfo{person}{Johannes
  Bader}, {and} \bibinfo{person}{Satish Chandra}.}
  \bibinfo{year}{2019}\natexlab{}.
\newblock \showarticletitle{Getafix: Learning to fix bugs automatically}.
\newblock \bibinfo{journal}{\emph{arXiv preprint arXiv:1902.06111}}
  (\bibinfo{year}{2019}).
\newblock


\bibitem[\protect\citeauthoryear{Smith, Barr, Le~Goues, and Brun}{Smith
  et~al\mbox{.}}{2015}]%
        {smith2015cure}
\bibfield{author}{\bibinfo{person}{Edward~K Smith}, \bibinfo{person}{Earl~T
  Barr}, \bibinfo{person}{Claire Le~Goues}, {and} \bibinfo{person}{Yuriy
  Brun}.} \bibinfo{year}{2015}\natexlab{}.
\newblock \showarticletitle{Is the cure worse than the disease? {Overfitting}
  in automated program repair}. In \bibinfo{booktitle}{\emph{Proceedings of the
  10th FSE}}. ACM, \bibinfo{pages}{532--543}.
\newblock


\bibitem[\protect\citeauthoryear{Soto and Le~Goues}{Soto and Le~Goues}{2018}]%
        {soto2018using}
\bibfield{author}{\bibinfo{person}{Mauricio Soto} {and} \bibinfo{person}{Claire
  Le~Goues}.} \bibinfo{year}{2018}\natexlab{}.
\newblock \showarticletitle{Using a probabilistic model to predict bug fixes}.
  In \bibinfo{booktitle}{\emph{Proceedings of the 25th SANER}}. {IEEE},
  \bibinfo{pages}{221--231}.
\newblock


\bibitem[\protect\citeauthoryear{Thomas, Nagappan, Blostein, and Hassan}{Thomas
  et~al\mbox{.}}{2013}]%
        {thomas2013impact}
\bibfield{author}{\bibinfo{person}{Stephen~W Thomas},
  \bibinfo{person}{Meiyappan Nagappan}, \bibinfo{person}{Dorothea Blostein},
  {and} \bibinfo{person}{Ahmed~E Hassan}.} \bibinfo{year}{2013}\natexlab{}.
\newblock \showarticletitle{The impact of classifier configuration and
  classifier combination on bug localization}.
\newblock \bibinfo{journal}{\emph{TSE}} \bibinfo{volume}{39},
  \bibinfo{number}{10} (\bibinfo{year}{2013}), \bibinfo{pages}{1427--1443}.
\newblock


\bibitem[\protect\citeauthoryear{Urli, Yu, Seinturier, and Monperrus}{Urli
  et~al\mbox{.}}{2018}]%
        {urli2018design}
\bibfield{author}{\bibinfo{person}{Simon Urli}, \bibinfo{person}{Zhongxing Yu},
  \bibinfo{person}{Lionel Seinturier}, {and} \bibinfo{person}{Martin
  Monperrus}.} \bibinfo{year}{2018}\natexlab{}.
\newblock \showarticletitle{How to design a program repair bot?: insights from
  the repairnator project}. In \bibinfo{booktitle}{\emph{Proceedings of the
  40th ICSE}}. ACM, \bibinfo{pages}{95--104}.
\newblock


\bibitem[\protect\citeauthoryear{Wang, Parnin, and Orso}{Wang
  et~al\mbox{.}}{2015}]%
        {wang2015evaluating}
\bibfield{author}{\bibinfo{person}{Qianqian Wang}, \bibinfo{person}{Chris
  Parnin}, {and} \bibinfo{person}{Alessandro Orso}.}
  \bibinfo{year}{2015}\natexlab{}.
\newblock \showarticletitle{Evaluating the usefulness of IR-based fault
  localization techniques}. In \bibinfo{booktitle}{\emph{Proceedings of the
  24th ISSTA}}. ACM, \bibinfo{pages}{1--11}.
\newblock


\bibitem[\protect\citeauthoryear{Wang and Lo}{Wang and Lo}{2014}]%
        {wang2014version}
\bibfield{author}{\bibinfo{person}{Shaowei Wang} {and} \bibinfo{person}{David
  Lo}.} \bibinfo{year}{2014}\natexlab{}.
\newblock \showarticletitle{Version {History}, {Similar} {Report}, and
  {Structure}: {Putting} {Them} {Together} for {Improved} {Bug}
  {Localization}}. In \bibinfo{booktitle}{\emph{Proceedings of the 22nd ICPC}}.
  ACM, \bibinfo{pages}{53--63}.
\newblock


\bibitem[\protect\citeauthoryear{Wei, Pei, Furia, Silva, Buchholz, Meyer, and
  Zeller}{Wei et~al\mbox{.}}{2010}]%
        {wei2010automated}
\bibfield{author}{\bibinfo{person}{Yi Wei}, \bibinfo{person}{Yu Pei},
  \bibinfo{person}{Carlo~A Furia}, \bibinfo{person}{Lucas~S Silva},
  \bibinfo{person}{Stefan Buchholz}, \bibinfo{person}{Bertrand Meyer}, {and}
  \bibinfo{person}{Andreas Zeller}.} \bibinfo{year}{2010}\natexlab{}.
\newblock \showarticletitle{Automated fixing of programs with contracts}. In
  \bibinfo{booktitle}{\emph{Proceedings of the 19th ISSTA}}. ACM,
  \bibinfo{pages}{61--72}.
\newblock


\bibitem[\protect\citeauthoryear{Weimer, Nguyen, {Le Goues}, and
  Forrest}{Weimer et~al\mbox{.}}{2009}]%
        {westley2009automatically}
\bibfield{author}{\bibinfo{person}{Westley Weimer}, \bibinfo{person}{ThanhVu
  Nguyen}, \bibinfo{person}{Claire {Le Goues}}, {and}
  \bibinfo{person}{Stephanie Forrest}.} \bibinfo{year}{2009}\natexlab{}.
\newblock \showarticletitle{Automatically finding patches using genetic
  programming}. In \bibinfo{booktitle}{\emph{Proceedings of the 31st ICSE}}.
  \bibinfo{publisher}{{IEEE}}, \bibinfo{pages}{364--374}.
\newblock


\bibitem[\protect\citeauthoryear{Wen, Chen, Wu, Hao, and Cheung}{Wen
  et~al\mbox{.}}{2018}]%
        {wen2018context}
\bibfield{author}{\bibinfo{person}{Ming Wen}, \bibinfo{person}{Junjie Chen},
  \bibinfo{person}{Rongxin Wu}, \bibinfo{person}{Dan Hao}, {and}
  \bibinfo{person}{Shing-Chi Cheung}.} \bibinfo{year}{2018}\natexlab{}.
\newblock \showarticletitle{Context-Aware Patch Generation for Better Automated
  Program Repair}. In \bibinfo{booktitle}{\emph{Proceedings of the 40th ICSE}}.
  \bibinfo{publisher}{IEEE/ACM}, \bibinfo{pages}{1--11}.
\newblock


\bibitem[\protect\citeauthoryear{Wen, Wu, and Cheung}{Wen
  et~al\mbox{.}}{2016}]%
        {wen2016locus}
\bibfield{author}{\bibinfo{person}{Ming Wen}, \bibinfo{person}{Rongxin Wu},
  {and} \bibinfo{person}{Shing-Chi Cheung}.} \bibinfo{year}{2016}\natexlab{}.
\newblock \showarticletitle{Locus: {Locating} bugs from software changes}. In
  \bibinfo{booktitle}{\emph{Proceedings of the 31st ASE}}. IEEE,
  \bibinfo{pages}{262--273}.
\newblock


\bibitem[\protect\citeauthoryear{White, Tufano, Martinez, Monperrus, and
  Poshyvanyk}{White et~al\mbox{.}}{2019}]%
        {white2019sorting}
\bibfield{author}{\bibinfo{person}{Martin White}, \bibinfo{person}{Michele
  Tufano}, \bibinfo{person}{Matias Martinez}, \bibinfo{person}{Martin
  Monperrus}, {and} \bibinfo{person}{Denys Poshyvanyk}.}
  \bibinfo{year}{2019}\natexlab{}.
\newblock \showarticletitle{Sorting and Transforming Program Repair Ingredients
  via Deep Learning Code Similarities}. In
  \bibinfo{booktitle}{\emph{Proceedings of the 26th SANER}}. {IEEE}.
\newblock


\bibitem[\protect\citeauthoryear{Wong, Xiong, Zhang, Hao, Zhang, and Mei}{Wong
  et~al\mbox{.}}{2014}]%
        {wong2014boosting}
\bibfield{author}{\bibinfo{person}{Chu-Pan Wong}, \bibinfo{person}{Yingfei
  Xiong}, \bibinfo{person}{Hongyu Zhang}, \bibinfo{person}{Dan Hao},
  \bibinfo{person}{Lu Zhang}, {and} \bibinfo{person}{Hong Mei}.}
  \bibinfo{year}{2014}\natexlab{}.
\newblock \showarticletitle{Boosting {Bug}-{Report}-{Oriented} {Fault}
  {Localization} with {Segmentation} and {Stack}-{Trace} {Analysis}}. In
  \bibinfo{booktitle}{\emph{Proceedings of the 30th ICSME}}. IEEE,
  \bibinfo{pages}{181--190}.
\newblock


\bibitem[\protect\citeauthoryear{Wong, Gao, Li, Abreu, and Wotawa}{Wong
  et~al\mbox{.}}{2016}]%
        {wong2016survey}
\bibfield{author}{\bibinfo{person}{W~Eric Wong}, \bibinfo{person}{Ruizhi Gao},
  \bibinfo{person}{Yihao Li}, \bibinfo{person}{Rui Abreu}, {and}
  \bibinfo{person}{Franz Wotawa}.} \bibinfo{year}{2016}\natexlab{}.
\newblock \showarticletitle{A survey on software fault localization}.
\newblock \bibinfo{journal}{\emph{TSE}} \bibinfo{volume}{42},
  \bibinfo{number}{8} (\bibinfo{year}{2016}), \bibinfo{pages}{707--740}.
\newblock


\bibitem[\protect\citeauthoryear{Xin and Reiss}{Xin and Reiss}{2017a}]%
        {xin2017identifying}
\bibfield{author}{\bibinfo{person}{Qi Xin} {and} \bibinfo{person}{Steven~P
  Reiss}.} \bibinfo{year}{2017}\natexlab{a}.
\newblock \showarticletitle{Identifying test-suite-overfitted patches through
  test case generation}. In \bibinfo{booktitle}{\emph{Proceedings of the 26th
  ISSTA}}. ACM, \bibinfo{pages}{226--236}.
\newblock


\bibitem[\protect\citeauthoryear{Xin and Reiss}{Xin and Reiss}{2017b}]%
        {xin2017leveraging}
\bibfield{author}{\bibinfo{person}{Qi Xin} {and} \bibinfo{person}{Steven~P
  Reiss}.} \bibinfo{year}{2017}\natexlab{b}.
\newblock \showarticletitle{Leveraging syntax-related code for automated
  program repair}. In \bibinfo{booktitle}{\emph{Proceedings of the 32nd ASE}}.
  {IEEE}/{ACM}, \bibinfo{pages}{660--670}.
\newblock


\bibitem[\protect\citeauthoryear{Xiong, Liu, Zeng, Zhang, and Huang}{Xiong
  et~al\mbox{.}}{2018}]%
        {xiong2018identifying}
\bibfield{author}{\bibinfo{person}{Yingfei Xiong}, \bibinfo{person}{Xinyuan
  Liu}, \bibinfo{person}{Muhan Zeng}, \bibinfo{person}{Lu Zhang}, {and}
  \bibinfo{person}{Gang Huang}.} \bibinfo{year}{2018}\natexlab{}.
\newblock \showarticletitle{Identifying patch correctness in test-based program
  repair}. In \bibinfo{booktitle}{\emph{Proceedings of the 40th ICSE}}. ACM,
  \bibinfo{pages}{789--799}.
\newblock


\bibitem[\protect\citeauthoryear{Xiong, Wang, Yan, Zhang, Han, Huang, and
  Zhang}{Xiong et~al\mbox{.}}{2017}]%
        {xiong2017precise}
\bibfield{author}{\bibinfo{person}{Yingfei Xiong}, \bibinfo{person}{Jie Wang},
  \bibinfo{person}{Runfa Yan}, \bibinfo{person}{Jiachen Zhang},
  \bibinfo{person}{Shi Han}, \bibinfo{person}{Gang Huang}, {and}
  \bibinfo{person}{Lu Zhang}.} \bibinfo{year}{2017}\natexlab{}.
\newblock \showarticletitle{Precise condition synthesis for program repair}. In
  \bibinfo{booktitle}{\emph{Proceedings of the 39th ICSE}}. IEEE/ACM,
  \bibinfo{pages}{416--426}.
\newblock


\bibitem[\protect\citeauthoryear{Xuan, Martinez, DeMarco, Clement, Marcote,
  Durieux, Le~Berre, and Monperrus}{Xuan et~al\mbox{.}}{2017}]%
        {xuan2017nopol}
\bibfield{author}{\bibinfo{person}{Jifeng Xuan}, \bibinfo{person}{Matias
  Martinez}, \bibinfo{person}{Favio DeMarco}, \bibinfo{person}{Maxime Clement},
  \bibinfo{person}{Sebastian~Lamelas Marcote}, \bibinfo{person}{Thomas
  Durieux}, \bibinfo{person}{Daniel Le~Berre}, {and} \bibinfo{person}{Martin
  Monperrus}.} \bibinfo{year}{2017}\natexlab{}.
\newblock \showarticletitle{Nopol: Automatic repair of conditional statement
  bugs in java programs}.
\newblock \bibinfo{journal}{\emph{TSE}} \bibinfo{volume}{43},
  \bibinfo{number}{1} (\bibinfo{year}{2017}), \bibinfo{pages}{34--55}.
\newblock


\bibitem[\protect\citeauthoryear{Yang, Zhikhartsev, Liu, and Tan}{Yang
  et~al\mbox{.}}{2017}]%
        {yang2017better}
\bibfield{author}{\bibinfo{person}{Jinqiu Yang}, \bibinfo{person}{Alexey
  Zhikhartsev}, \bibinfo{person}{Yuefei Liu}, {and} \bibinfo{person}{Lin Tan}.}
  \bibinfo{year}{2017}\natexlab{}.
\newblock \showarticletitle{Better test cases for better automated program
  repair}. In \bibinfo{booktitle}{\emph{Proceedings of the 11th FSE}}. ACM,
  \bibinfo{pages}{831--841}.
\newblock


\bibitem[\protect\citeauthoryear{Yoo and Harman}{Yoo and Harman}{2012}]%
        {yoo2012regression}
\bibfield{author}{\bibinfo{person}{Shin Yoo} {and} \bibinfo{person}{Mark
  Harman}.} \bibinfo{year}{2012}\natexlab{}.
\newblock \showarticletitle{Regression testing minimization, selection and
  prioritization: a survey}.
\newblock \bibinfo{journal}{\emph{STVR}} \bibinfo{volume}{22},
  \bibinfo{number}{2} (\bibinfo{year}{2012}), \bibinfo{pages}{67--120}.
\newblock


\bibitem[\protect\citeauthoryear{Youm, Ahn, and Lee}{Youm
  et~al\mbox{.}}{2017}]%
        {youm2017improved}
\bibfield{author}{\bibinfo{person}{Klaus~Changsun Youm}, \bibinfo{person}{June
  Ahn}, {and} \bibinfo{person}{Eunseok Lee}.} \bibinfo{year}{2017}\natexlab{}.
\newblock \showarticletitle{Improved bug localization based on code change
  histories and bug reports}.
\newblock \bibinfo{journal}{\emph{IST}}  \bibinfo{volume}{82}
  (\bibinfo{year}{2017}), \bibinfo{pages}{177--192}.
\newblock


\bibitem[\protect\citeauthoryear{Yu, Martinez, Danglot, Durieux, and
  Monperrus}{Yu et~al\mbox{.}}{2017}]%
        {yu2017test}
\bibfield{author}{\bibinfo{person}{Zhongxing Yu}, \bibinfo{person}{Matias
  Martinez}, \bibinfo{person}{Benjamin Danglot}, \bibinfo{person}{Thomas
  Durieux}, {and} \bibinfo{person}{Martin Monperrus}.}
  \bibinfo{year}{2017}\natexlab{}.
\newblock \showarticletitle{Test case generation for program repair: A study of
  feasibility and effectiveness}.
\newblock \bibinfo{journal}{\emph{arXiv preprint arXiv:1703.00198}}
  (\bibinfo{year}{2017}).
\newblock


\bibitem[\protect\citeauthoryear{Yu, Martinez, Danglot, Durieux, and
  Monperrus}{Yu et~al\mbox{.}}{2018}]%
        {yu2018alleviating}
\bibfield{author}{\bibinfo{person}{Zhongxing Yu}, \bibinfo{person}{Matias
  Martinez}, \bibinfo{person}{Benjamin Danglot}, \bibinfo{person}{Thomas
  Durieux}, {and} \bibinfo{person}{Martin Monperrus}.}
  \bibinfo{year}{2018}\natexlab{}.
\newblock \showarticletitle{Alleviating patch overfitting with automatic test
  generation: a study of feasibility and effectiveness for the Nopol repair
  system}.
\newblock \bibinfo{journal}{\emph{EMSE Journal}} (\bibinfo{year}{2018}),
  \bibinfo{pages}{1--35}.
\newblock


\bibitem[\protect\citeauthoryear{Zhou, Zhang, and Lo}{Zhou
  et~al\mbox{.}}{2012}]%
        {zhou2012should}
\bibfield{author}{\bibinfo{person}{Jian Zhou}, \bibinfo{person}{Hongyu Zhang},
  {and} \bibinfo{person}{David Lo}.} \bibinfo{year}{2012}\natexlab{}.
\newblock \showarticletitle{Where should the bugs be fixed? more accurate
  information retrieval-based bug localization based on bug reports}. In
  \bibinfo{booktitle}{\emph{Proceedings of the 34th ICSE}}. IEEE,
  \bibinfo{pages}{14--24}.
\newblock


\bibitem[\protect\citeauthoryear{Zimmermann, Premraj, Bettenburg, Just,
  Schroter, and Weiss}{Zimmermann et~al\mbox{.}}{2010}]%
        {zimmermann2010makes}
\bibfield{author}{\bibinfo{person}{Thomas Zimmermann}, \bibinfo{person}{Rahul
  Premraj}, \bibinfo{person}{Nicolas Bettenburg}, \bibinfo{person}{Sascha
  Just}, \bibinfo{person}{Adrian Schroter}, {and} \bibinfo{person}{Cathrin
  Weiss}.} \bibinfo{year}{2010}\natexlab{}.
\newblock \showarticletitle{What makes a good bug report?}
\newblock \bibinfo{journal}{\emph{TSE}} \bibinfo{volume}{36},
  \bibinfo{number}{5} (\bibinfo{year}{2010}), \bibinfo{pages}{618--643}.
\newblock


\end{thebibliography}

\end{document}